\documentclass[sigconf]{acmart}
\AtBeginDocument{%
  }
\usepackage{xspace}
\usepackage{subcaption}
\usepackage{CJKutf8}
\usepackage[russian, english]{babel}
\usepackage{enumitem}
\usepackage{tabularx}
\usepackage{graphicx}
\usepackage{array}
\usepackage{color}
\usepackage{booktabs} 
\usepackage{xcolor} 
\usepackage{colortbl} 
\usepackage{url} 

\newlist{tightitem}{itemize}{1}
\setlist[tightitem]{nosep, leftmargin=1.5em, label=--}

\usepackage[capitalise]{cleveref}
\crefformat{section}{\S#2#1#3}
\crefname{figure}{Figure}{Figures}
\crefname{appendix}{Appendix}{Appendices}
\crefname{table}{Table}{Tables}
\crefname{algorithm}{Algorithm}{Algorithms}
\crefname{listing}{Listing}{Listings}
\crefname{theorem}{Theorem}{Theorems}
\crefname{thm}{Theorem}{Theorems}
\crefname{lemma}{Lemma}{Lemmata}
\crefname{equation}{Eqt.}{Eqts.}
\crefformat{Grammar}{Grammar #1}
\crefformat{RQ}{RQ #1}

\newcommand{\etal}{\textit{et al.}\xspace}
\newcommand{\myparagraph}[1]{\textbf{#1}}
\newcommand{\numberofcomments}{{9.14 billion}\xspace}
\newcommand{\numberofissuecomments}{{1.7 billion}\xspace}
\newcommand{\numberofrepos}{62{,}500\xspace}
\newcommand{\numberofpl}{five\xspace}
\newcommand{\numberofnl}{30\xspace}

\copyrightyear{2026}
\acmYear{2026}
\setcopyright{cc}
\setcctype{by}
\acmConference[ICSE '26]{2026 IEEE/ACM 48th International Conference on Software Engineering}{April 12--18, 2026}{Rio de Janeiro, Brazil}
\acmBooktitle{2026 IEEE/ACM 48th International Conference on Software Engineering (ICSE '26), April 12--18, 2026, Rio de Janeiro, Brazil}
\acmPrice{}
\acmDOI{10.1145/3744916.3787766}
\acmISBN{979-8-4007-2025-3/2026/04}

\begin{document}

\title{``Write in English, Nobody Understands Your Language Here'': \\A Study of  Non-English Trends in Open-Source Repositories\footnotemark}

\author{Masudul Hasan Masud Bhuiyan}
\email{masudul.bhuiyan@cispa.de}
\orcid{0000-0002-7090-4334}
\affiliation{%
  \institution{CISPA Helmholtz Center for Information Security}
  \city{Saarbrücken}
  \country{Germany}
}

\author{Manish Kumar Bala Kumar}
\email{manish.bala-kumar@cispa.de}
\orcid{0009-0000-8964-6542}
\affiliation{%
  \institution{CISPA Helmholtz Center for Information Security}
  \city{Saarbrücken}
  \country{Germany}
}

\author{Cristian-Alexandru Staicu}
\email{cris.staicu@gmail.com}
\orcid{0000-0002-6542-2226}
\affiliation{%
  \institution{CISPA Helmholtz Center for Information Security}
  \city{Saarbrücken}
  \country{Germany}
}

\renewcommand{\shorttitle}{``Write in English, Nobody Understands Your Language Here'': \\A Study of  Non-English Trends in Open-Source Repositories}

\begin{abstract}

The open-source software (OSS) community has historically been dominated by English as the primary language for code, documentation, and developer interactions. However, with growing global participation and better support for non-Latin scripts through standards like Unicode, OSS is gradually becoming more multilingual. This study investigates the extent to which OSS is becoming more multilingual, analyzing \numberofcomments GitHub issues, pull requests, and discussions, and \numberofrepos repositories across \numberofpl programming languages and \numberofnl natural languages, covering the period from 2015 to 2025.
We examine six research questions to track changes in language use across communication, code, and documentation. We find that multilingual participation has steadily increased, especially in Korean, Chinese, and Russian. This growth appears not only in issues and discussions but also in code comments, string literals, and documentation files.
While this shift reflects greater inclusivity and language diversity in OSS, it also creates language tension. The ability to express oneself in a native language can clash with shared norms around English use, especially in collaborative settings. Non-English or multilingual projects tend to receive less visibility and participation, suggesting that language remains both a resource and a barrier, shaping who gets heard, who contributes, and how open collaboration unfolds. 
\end{abstract}


\maketitle
\footnotetext{The title of the paper is inspired by this conversation in an open-source repository: \url{https://github.com/dariowouters/ts-fmod-plugin/issues/34}}
\section{Introduction}
The open-source software (OSS) ecosystem has been a cornerstone of modern software development, fostering collaboration among developers worldwide. Historically, English has served as the lingua franca for OSS, dominating codebases, documentation, and community interactions. Platforms like Stack Overflow require all questions and answers to be in English~\cite{stackoverflowQuestionLanguage}, and most GitHub discussions also occur in English. This dominance is rooted in the early history of computing, when research and development were led by the United States and United Kingdom, and early programming languages were designed in English~\cite{mahoney2005histories}. As a result, developers globally adopted English to write code and use software tools, regardless of their native language.

Over the years, software tools have improved support for diverse writing systems. The release of Unicode 3.0 in 1999 enabled more reliable encoding of non-Latin scripts. Today, scripts such as Chinese, Arabic, Cyrillic, and Bengali are widely used across digital writing, websites~\cite{bhuiyan2025not}, and software projects. Web technologies followed this shift, and modern browsers and development tools now support many scripts. Ebbertz \etal showed that English accounted for 80\% of web content in 1998, dropping to 56\% by 2002~\cite{ebbertz2002internet}. Recent estimates place this share between 20
At the same time, the global software development landscape has shifted. Nearly a quarter of the open-source community reports limited English proficiency\footnote{\url{https://opensourcesurvey.org/2017/}}. GitHub growth data shows that some of the fastest-growing contributor communities are in non-English-dominant countries, including China, Brazil, India, and Russia~\cite{githubGlobalDistribution2022, githubOctoverseLeads2024, githubOctoverseState2023}. These changes raise important questions about the role of English in online collaboration. 

Prior research provides limited insight into the linguistic dynamics of open-source software development. Pawelka \etal ~\cite{pawelka2015code} analyzed natural language use in comments and identifiers across 23 software projects. They found that while industry projects contain some non-English text, open-source projects contain almost none, pointing to a strong English-language bias in public code repositories. To the best of our knowledge, this remains the only study that has examined the presence of non-English languages in code repositories. However, their study does not explore whether this trend has evolved over time.

Beyond the single study on non-English content in code, others have extensively studied how language affects collaboration and participation, especially for newcomers. Yoseph \etal identified language-related challenges as the third most commonly reported bias faced by contributors in open-source communities~\cite{alebachew2025pageexaminingdeveloperperception}. Steinmacher \etal reported that limited English proficiency is one of the major barriers preventing newcomers from communicating or participating in discussions in open-source projects~\cite{steinmacher2015social}. Similarly, Balali \etal found that differences in language fluency can negatively affect mentoring relationships, making onboarding more difficult both socially and technically~\cite{balali2018newcomers}. However, these findings are based on limited samples and also do not track longitudinal changes or large-scale usage across repositories. 
GitHub also publishes annual Octoverse statistics~\cite{githubOctoverseState2023, githubOctoverseLeads2024} that track demographics, programming language trends, and regional growth. While useful, these reports do not analyze the natural languages used in issues, pull requests, documentation, or code. Demographics can overlap with linguistic patterns, but the relationship is not linear: multiple regions share the same language (e.g., Arabic, Spanish), and many countries have several official languages (e.g., India). As a result, demographic counts alone do not indicate the languages developers actually use on the platform.

Despite these insights, there is still a gap in the literature on how language use is evolving in open-source projects over time. A key question remains: How is it changing across code, discussion, and documentation?
In this paper, we present the first large-scale empirical study of multilingualism in open source software development. We analyze a dataset of \numberofcomments GitHub messages, including issues, comments, pull requests, reviews, and discussions, collected from 2015 to May 2025. To examine language use in code, we also study \numberofrepos repositories over the same period. These repositories span five widely used programming languages: JavaScript, Python, TypeScript, Java, and C\#. Using language detection and classification tools, we identify the presence of \numberofnl natural languages across different parts of open source collaboration, including code elements such as identifiers and string literals, documentation, and user-generated messages. In particular, our study answers the following six research questions:
\begin{itemize}[leftmargin=5pt]
    \item \textit{RQ1: Is developer discussion in open-source becoming more multilingual?} Our results show that non-English content in issues and pull requests grew by 164 percent between 2015 and 2025.

    \item \textit{RQ2: Which natural languages are most prevalent in open source interactions?}
    Korean usage increased by 1706 percent, and Chinese by 120 percent. Along with Japanese, Russian, and Vietnamese, they lead non-English content on GitHub.
    
    \item \textit{RQ3: Is the code itself, such as string literals and identifiers, becoming more multilingual?}
    Yes, mostly in comments and literals. Non-English comments rose from 3.6 to 11.9 percent, and literals rose from 3.2 to 9.6 percent. Identifiers stayed mostly English.
    
    \item \textit{RQ4: Are there significant differences across programming languages?}
    Yes. Java, Python, and JavaScript show more non-English use than C\# and TypeScript. Java has the highest share of multilingual comments and literals.
    
    \item \textit{RQ5: Are there repositories where developers regularly use multiple languages? Are there signs of language-based friction or coordination issues?}
    We found 193,000 repositories using multiple languages. But language often creates tension for non-English participants.

    \item \textit{RQ6: Does non-English content negatively affect software development practices?}
    Yes. English-dominant projects tend to receive more visibility, collaboration, and participation.
\end{itemize}

\section{Related Work}
Prior work on multilingualism and open-source development has examined natural language in code, the impact of language barriers on collaboration~\cite{steinmacher2019overcoming, prikladnicki2003global}, communication practices on platforms like GitHub~\cite{denny2021designing, reestman2019native}, and how language shapes inclusivity~\cite{miller2022did, sultana2024assessinginfluencetoxicgender, landscapetoxic}. Researchers have also studied user behavior in forums~\cite{hellman2022characterizing}, social and collaborative structures~\cite{Avelino_2016}, and diversity in developer communities~\cite{genderrepresentation, terrell2016gender}. These works highlight barriers and individual practices, but few analyze multilingualism at scale or over time. Most are qualitative or focus on isolated aspects. In contrast, our study offers a longitudinal, large-scale analysis of multilingual trends across code, documentation, and discussion.

\myparagraph{Language Barriers in Collaboration:}
English proficiency is a recurrent barrier for contributors, especially newcomers~\cite{feng2025multifaceted, fatima2025developer, zhao2004user, noll2011global, steinmacher2015systematic, steinmacher2019overcoming, prikladnicki2003global}. Steinmacher \etal~\cite{steinmacher2015systematic} identified language as a consistent challenge for new contributors. Balali \etal~\cite{balali2018newcomers} found that differences in English fluency hinder mentoring and communication. In global software teams, Noll \etal~\cite{noll2011global} showed that non-native speakers face difficulties in meetings, documentation, and cross-cultural understanding, affecting coordination and productivity. These studies focus on onboarding or team settings, whereas our work examines multilingual trends in open source over time.

\myparagraph{Language in Developer Discussions:}
Research has explored how language shapes communication on platforms such as GitHub and Stack Overflow~\cite{kavaler2017perceived, guo2018non, becker2019parlez, becker2018fix, bregolin2022communication, ortu2015would, denny2021designing, reestman2019native}. Kavaler \etal~\cite{kavaler2017perceived} linked higher linguistic complexity in issues to slower resolution. Guo \etal~\cite{guo2018non} showed that non-native speakers struggle with documentation and informal expressions. Bregolin \etal~\cite{bregolin2022communication} found that native-language Stack Overflow sites increase participation without harming the English site. Becker \etal~\cite{becker2018fix} showed that technical or unnatural English in error messages challenges both non-native and native speakers. These works target specific interactions rather than long-term multilingual trends.

\myparagraph{Language and Toxicity in Open Source:}
Language is closely tied to communication dynamics in OSS. Miller \etal~\cite{miller2022did} showed that ambiguity and non-responsiveness can escalate into negative exchanges. Sultana \etal~\cite{sultana2024assessinginfluencetoxicgender} linked toxic communication to reduced diversity. Sarker \etal~\cite{ToxiSpanSE, landscapetoxic} developed methods to detect toxicity in code reviews, showing how linguistic cues can signal exclusion.
\begin{figure*}
    \begin{center}
      \includegraphics[width=.8\linewidth, keepaspectratio]{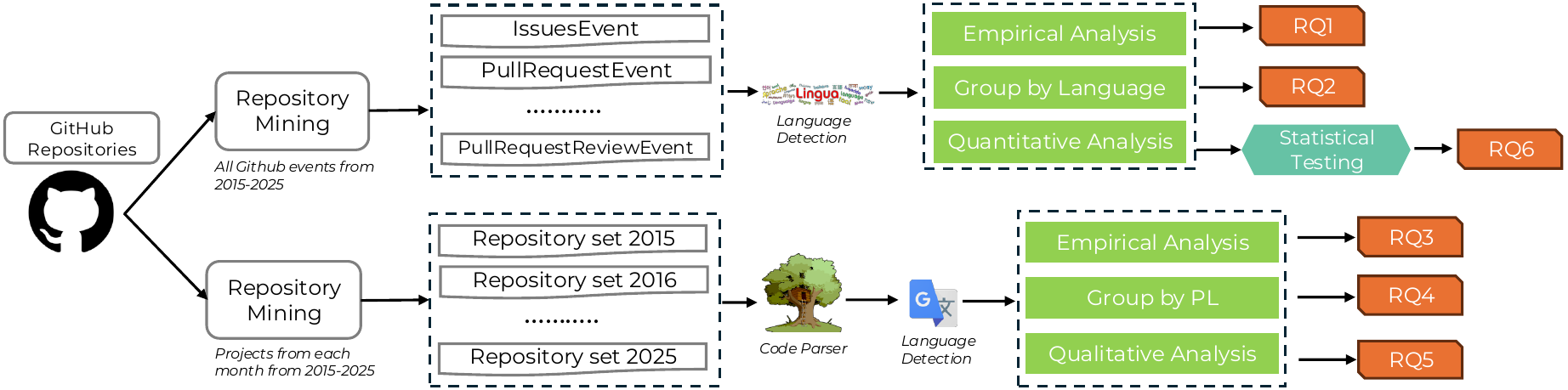}
    \end{center}
    \vspace{-3mm}
    \caption{Overview of our methodology: data collection, language detection, and analysis across code and discussions.}
    \label{fig:methodology}
    \vspace{-3mm}
\end{figure*}

\myparagraph{Platform Reports on Developer Activity:}
GitHub’s annual Octoverse report summarizes platform activity, highlighting trends in programming languages, repository growth, and developer demographics~\cite{githubOctoverseState2023, githubOctoverseLeads2024}. Stack Overflow’s annual survey similarly covers demographics and technologies~\cite{stackoverflowStackOverflow}. While these reports provide valuable statistics, they emphasize technical dimensions rather than the linguistic composition of developer activity. They do not examine language use in discussions or code, overlooking multilingual factors that shape open-source collaboration. For example, Octoverse tracks regional growth in countries such as India, Indonesia, and Brazil, but does not assess whether this growth corresponds to greater linguistic diversity in repositories or discussions. Our study addresses this gap through large-scale, language-aware measurements of communication and code that go beyond aggregate reports.

\section{Study Design and Methodology}
\label{sec:methodology_overview}

To examine the role of multilingualism in open-source software development, we analyzed GitHub activity from January~2015 to May~2025. Open-source platforms contain multiple forms of natural language content. Some elements facilitate communication among developers, such as issue descriptions, pull request discussions, and review comments, while others, like commit messages, are transactional and not central to this study. Because the available data sources differ in scope and scalability, our approach combines two complementary analysis pipelines, each aligned with specific research questions. 
The first pipeline (used for \textbf{RQ1}, \textbf{RQ2}, and \textbf{RQ6}) focuses on communication-related activity drawn from \textsc{GHArchive}. This dataset provides large-scale event metadata, including issues, pull requests, and associated comments. The second pipeline (used for \textbf{RQ3}, \textbf{RQ4}, and \textbf{RQ5}) targets source code and documentation, analyzing identifiers, literals, and code comments. Since \textsc{GHArchive} does not include source code content, we use the GitHub REST API to retrieve code patches. API access is rate-limited and less scalable, so this dataset covers a representative, temporally balanced sample of repositories rather than the full population. \Cref{fig:methodology} summarizes both pipelines and the overall data flow.
For RQ6, we also performed statistical testing to assess whether the observed differences across language categories are significant. 
Given the skewed nature of repository activity data, we used non-parametric methods suitable for comparing multiple groups. 
We report overall group differences and pairwise comparisons where relevant.

All analyses were performed on a dedicated Linux server equipped with an AMD~EPYC~7H12 processor (128~physical cores, 256~threads), 2~TB~of~RAM, and 60~TB~of~storage. 
Processing the entire dataset required several weeks and included the classification of over \numberofcomments GitHub messages and the analysis of \numberofrepos sampled code patches. 
The pipeline covered metadata extraction, text preprocessing, and natural language identification across both communication and code elements.
In the following sections, we present the detailed methodology and results for each research question. 
Each section explains the specific subset of data used, the metrics derived from it, and the statistical analyses applied, followed by quantitative findings and interpretation.
\section{RQ1: Is open source development becoming more multilingual?}
\label{sec:rq1}
This research question examines whether open-source development has become more multilingual over time and how the linguistic composition of GitHub communication has changed.

\myparagraph{Data Collection.}
We analyzed all available GitHub discussions from January 2015 to May 2025, including issues, pull requests, and their associated comments and reviews. This resulted in a dataset of approximately \numberofcomments{} messages, drawn from over 476 million repositories and spanning contributions from 78.01 million developers. The data was collected from GHArchive, a public record of GitHub activity that tracks user interactions and event metadata over time. We used the May 2025 release of GHArchive and processed a total of 4.3~TB of compressed JSON data.
To extract full message content, we parsed it directly from GHArchive event records, which include both the message text and relevant metadata for each interaction. We focused on five types of user interactions where natural language is commonly used: creating and commenting on issues, opening and reviewing pull requests, and posting comments during code reviews. These represent the main communication channels used by developers on GitHub.

\begin{figure}
    \centering
    \begin{subfigure}[t]{0.48\linewidth}
        \includegraphics[width=\linewidth]{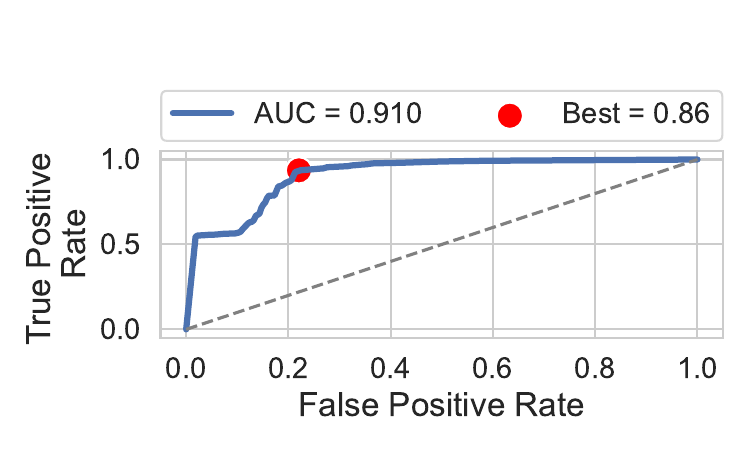}
        \caption{{Lingua} ROC curve.}
        \label{fig:lingua_auc}
    \end{subfigure}
    \begin{subfigure}[t]{0.48\linewidth}
        \includegraphics[width=\linewidth]{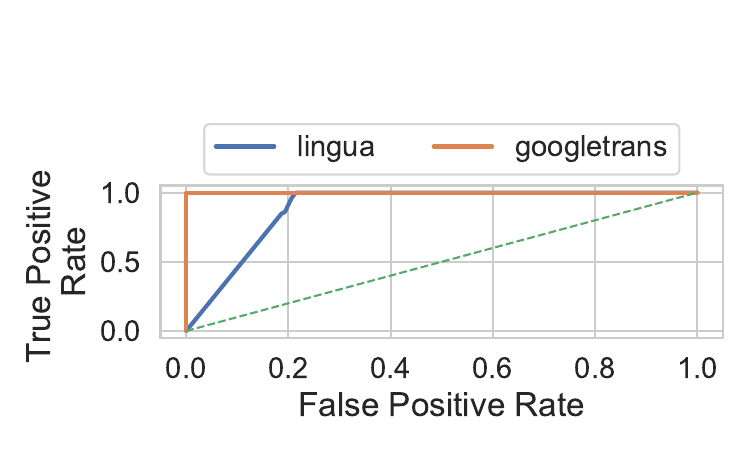}
        \caption{{Lingua} vs Google Translator.}
        \label{fig:lingua_google_comparison}
    \end{subfigure}

    \vspace{-2mm}
    \caption{Evaluation of language detection models: (a) ROC curve and threshold selection for \texttt{Lingua}; (b) performance comparison with Google Translator on short code texts.}
    \label{fig:lingua_evaluation}
    \vspace{-4mm}
\end{figure}

\myparagraph{Language Detection.}
We used the \texttt{Lingua}~\cite{lingua} language identification library (v2.1.1) to classify the natural language of each message. After filtering for relevant message types, we retained \numberofissuecomments{} data points containing natural language text. 
Language detection on GitHub data is challenging because many messages are short, contain typos, or include unformatted code, which can confuse classifiers. 
To reduce these errors, we used Lingua’s strictest detection mode and excluded all messages with low-confidence results. We evaluated the classifier’s performance using a manually annotated sample of messages and computed the receiver operating characteristic (ROC) curve shown in \cref{fig:lingua_auc}. 
The model achieved an area under the curve (AUC) of~0.91, with the optimal confidence threshold at~0.86. 
To remain conservative and minimize false positives, we set our cutoff slightly higher, at~0.9. 
Messages below this confidence level were excluded from the analysis and from the percentage calculations. 
Lingua supports 75~languages, and we accepted any message with a high-confidence match, while focusing our analysis on non-Latin scripts and non-English text.

\begin{figure}
\begin{center}
  \includegraphics[width=.9\linewidth, keepaspectratio]{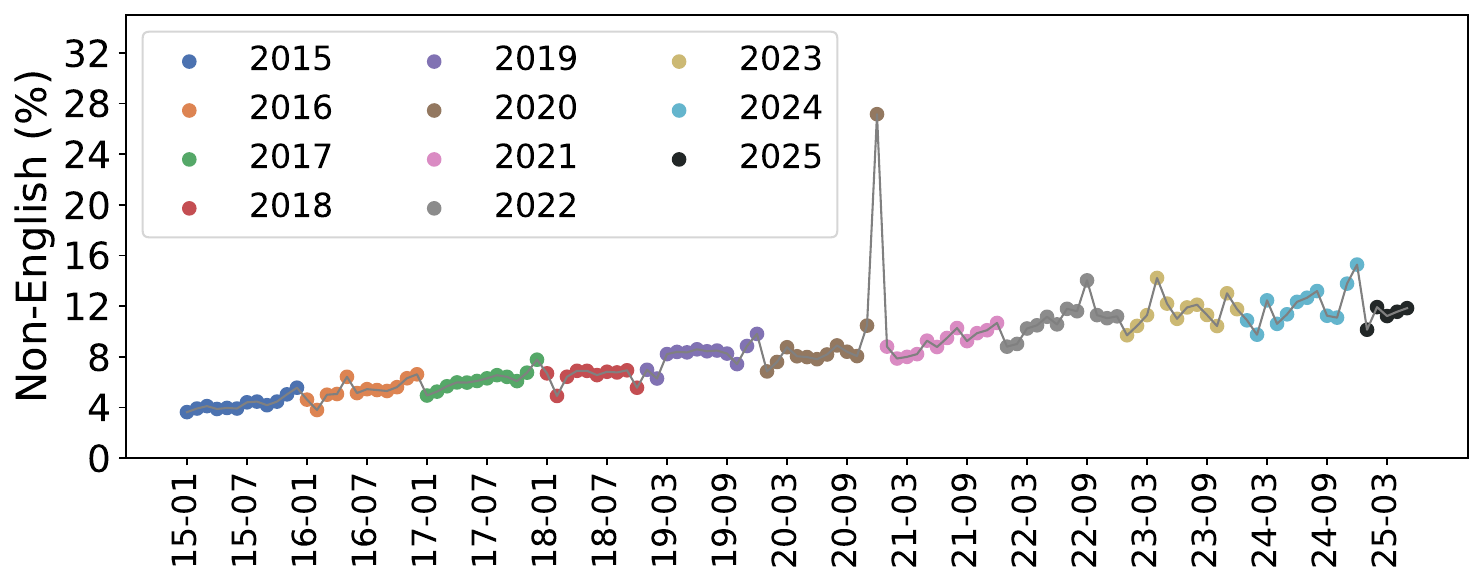}
\end{center}
\vspace{-3mm}
\caption{Monthly percentage of non-English content in GitHub discussions (issues, pull requests) from 2015 to 2025, showing a consistent upward trend and increased multilingual participation in open-source development.}
\label{fig:non_english_timeline}
\vspace{-5mm}
\end{figure}

\myparagraph{Results.}
As shown in \cref{fig:non_english_timeline}, the proportion of non-English content increased consistently over this period, rising from an average of 4.2\% in~2015 to 11.3\% in~2025, representing a 164\% relative increase over the decade. 
Monthly data show a steady upward trajectory, with notable acceleration after~2019. 
In~2019, the average proportion of non-English content was 8.2\%, increasing to 9.9\% in~2020, marking a structural shift in participation dynamics. 
From~2022 onward, the percentage of non-English messages remained above~10\% annually, signaling a lasting change in the linguistic makeup of open-source communities.

\myparagraph{Observations.}
A few notable patterns emerge. First, there is a sharp spike in December 2020, where non-English content briefly reaches 27.1\%. Upon manual inspection of the underlying data, we found that this spike corresponded to a sudden surge in issues containing Chinese text, posted across multiple repositories. These issues accounted for a large volume of content in that month. However, none of the issue links are currently accessible, suggesting that the repositories or issues may have been deleted, made private, or removed from GitHub.
The content of these issues points to a coordinated spam attack related to promoting WeChat groups. Many messages contained phrases such as \begin{CJK}{UTF8}{gbsn}``澳洲幸运10信誉公众号微信群''\end{CJK} (``Australia Lucky 10 trusted public WeChat group''), \begin{CJK}{UTF8}{gbsn}``极速赛车微信群游戏''\end{CJK} (``Speed racing WeChat group game''), and \begin{CJK}{UTF8}{gbsn}``比较信誉极速赛车微信公众号平台''\end{CJK} (``Compare trusted speed racing WeChat public account platforms''). The repeated references to WeChat, group promotions, and gambling-like keywords suggest the posts were meant to advertise or exploit access to Chinese-language WeChat groups.

We observed a similar anomaly on May 1, 2016, when a large number of issues containing Russian text appeared across several repositories. These entries likewise drove a temporary increase in non-English content. As with the 2020 spike, the original links to these issues are no longer available. While both events appear to be isolated, they reveal that even attackers assume some level of language diversity and inclusivity. They adapt their attacks to the target audience’s language, leveraging localized communication to increase effectiveness.
Second, the persistence of double-digit non-English percentages from 2022 onward suggests a real decline in English dominance. For example, in January 2015, 3.6\% of issue comments were non-English. By May 2025, that number rose to 11.8\%. It is important to note that these values are percentages, not absolute counts. GitHub activity has grown significantly over time, but the use of percentages accounts for that underlying growth.

\section{RQ2: Which natural languages are most prevalent in open-source interactions?}
\label{sec:rq2}

This research question identifies the most commonly used non-English natural languages in open-source discussions on GitHub and how their prevalence has evolved over time. 
We use the same dataset and methodology described in \cref{sec:rq1}, based on communication data from \textsc{GHArchive} between January~2015 and May~2025.

\begin{figure}
\begin{center}
  \includegraphics[width=.8\linewidth, keepaspectratio]{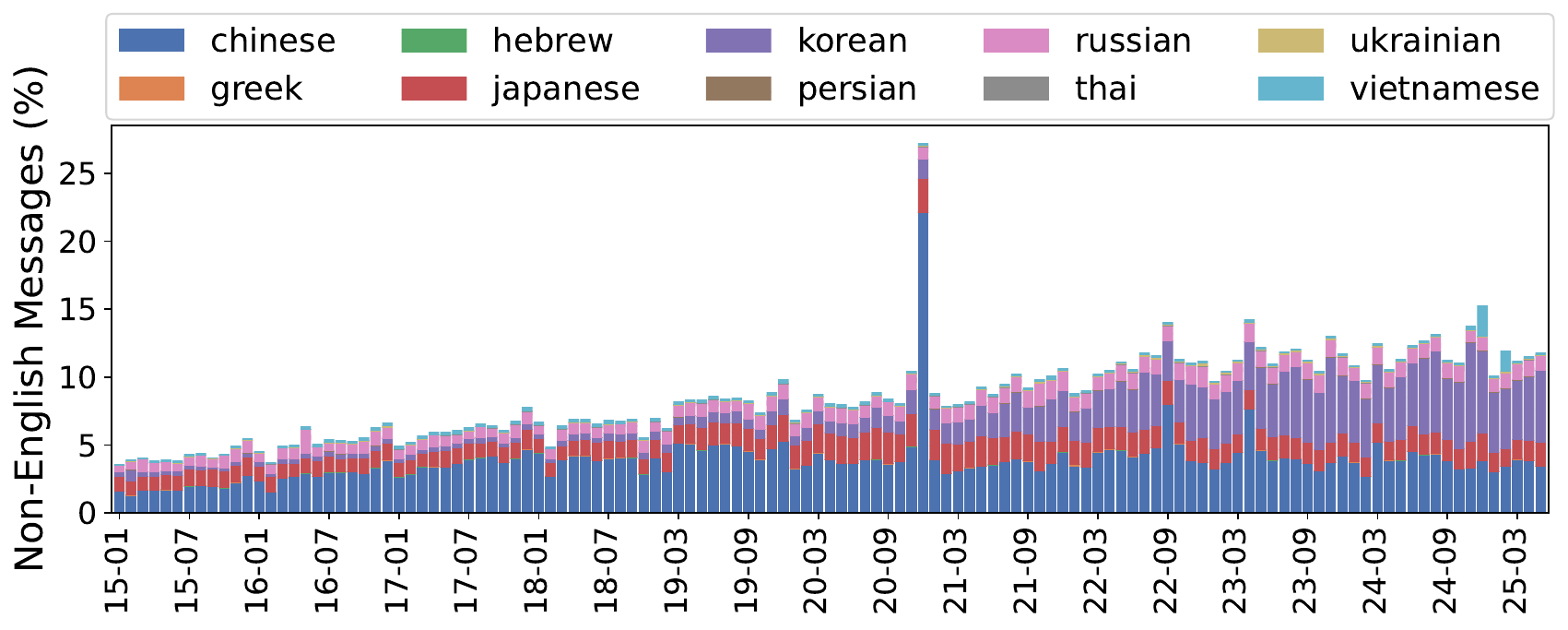}
\end{center}
\vspace{-3mm}
\caption{Monthly distribution of non-English natural languages in GitHub discussions from 2015 to 2025, highlighting the sustained presence of Chinese, Japanese, Russian, Korean, and Vietnamese, and the growing participation in Korean, Persian, and Vietnamese over time.}
\label{fig:language_distribution_rq3}
\vspace{-5mm}
\end{figure}

\myparagraph{Results.}
\Cref{fig:language_distribution_rq3} shows the distribution of natural languages across all GitHub messages over the ten-year period. 
The top five languages by average share across the 10-year period are Chinese (3.8\%), Korean (1.9\%), Japanese (1.5\%), Russian (1.0\%), and Vietnamese (0.18\%). Other languages such as Ukrainian (0.06\%), Thai (0.02\%), Persian (0.02\%), Hebrew (0.01\%), and Greek (0.01\%) contributed minimally on average.
Chinese consistently appears as the most dominant non-English language throughout the decade, growing from 1.6\% in January 2015 to 3.4\% in May 2025. Its influence peaked in December 2020, when it spiked to 22.1\% of all messages, likely contributing to the multilingual surge observed in \cref{sec:rq1}. Korean shows the most significant growth, rising from 0.3\% in early 2015 to 5.2\% in May 2025, with a peak of 7.3\% in November 2024. Japanese usage remains steady, fluctuating around 1.1\% to 1.7\%, while Russian gradually increases from 0.5\% to 1.1\%. Vietnamese has a low average but shows occasional spikes, reaching 2.2\% in December 2024.

\myparagraph{Observations.}These results highlight three key observations. First, the top five non-English languages, Chinese, Korean, Japanese, Russian, and Vietnamese, account for the vast majority of multilingual content on GitHub. Their consistent presence over the 10-year period suggests that these languages correspond to stable and active developer communities, rather than sporadic or event-driven participation.

Second, the rapid growth of Korean messages, particularly after 2021, signals a rising open-source contributor base in South Korea. Korean-language content increased from 0.3\% in 2015 to 5.2\% in 2025, despite South Korea representing less than 1\% of the global population. This contrast highlights an unusually high level of open-source engagement relative to country size.
National initiatives have likely played a role: for example, the Korean government's \textit{Open Source Software Invigoration Plan} (2014) and the 2020 amendment to the \textit{Software Promotion Act} mandated that publicly funded software projects release their code under open-source licenses~\cite{eu2022koreaoss}. There are also active Korean open-source groups, such as the Linux Foundation's OpenChain Korea Work Group\footnote{\url{https://github.com/OpenChain-Project/OpenChain-KWG}}, and prominent projects like SeoulTech's \texttt{open-data-seoul}\footnote{\url{https://github.com/SeoulTech/open-data-seoul}}, which further illustrate a growing open-source ecosystem.

Third, although the average percentages for Thai, Vietnamese, and Persian remain below 0.2\%, all three show clear upward trends. Vietnamese usage increases steadily, while Persian reaches 0.13\% by early 2025. The rise in Persian-language content is especially notable given the context of international sanctions and access restrictions facing Iranian developers. These growing contributions suggest persistent effort from developers in regions often underrepresented in global software ecosystems.
\section{RQ3: Is source code becoming more multilingual?}
\label{sec:rq3}

We now turn to the source code to examine whether developer-written elements—such as comments, literals, and identifiers—show increasing use of non-English text over time.

\myparagraph{Temporal Sampling.}
To capture temporal changes in the use of natural language within code, we sampled 385~patches per month from January~2015 to May~2025. 
This number was derived from an estimated 1.12~billion public GitHub contributions~\cite{octoverse2025}, providing a 95\% confidence level with a 5\% margin of error. 
We first filtered for patches that introduced at least one new file within each push event to ensure sufficient syntactic context, as new files typically contain complete function definitions, class declarations, comments, and string literals—the main elements where natural language appears. 
Parsing partial patches, in contrast, is difficult and often yields incomplete or noisy syntax. 
To remove trivial or auto-generated changes, we applied a 500-character minimum threshold. 
Such filtering helps retain meaningful development activity and sufficient semantic context; prior mining studies likewise exclude very small commits or diffs for this reason~\cite{kondo2024empirical, hindle2008large}. 
We then randomly selected 385~patches per month (one per repository), resulting in a total of 48{,}125~patches. 
We chose this approach because partial patches are hard to parse and rarely provide enough syntactic context. New files, in contrast, usually contain full function definitions, class declarations, comments, and string literals, the main places where natural language appears.
The same strategy was applied to documentation files to analyze multilingual README content.

\myparagraph{Code Parsing.}
We selected the top five programming languages on GitHub~\cite{githubOctoverseLeads2024}: JavaScript, Python, Java, TypeScript, and C\#, to ensure broad coverage across dominant ecosystems. 
Each source file was parsed using \texttt{tree-sitter}~\cite{treesitterIntroductionTreesitter} to extract and tokenize the following elements based on each language’s grammar:
The parser extracted and tokenized the following language elements based on each language’s grammar:
\begin{tightitem}
\item \textbf{Identifiers:} user-defined variable and symbol names, excluding function and class declarations.
\item \textbf{Function and Class Names:} identifiers used to declare functions, methods, and classes.
\item \textbf{Literals:} string literals excluding minified or template-embedded content.
\item \textbf{Comments:} inline, block, and documentation comments (e.g., Python docstrings~\cite{pythonDocstringConventions} or JavaDoc~\cite{oracleJavadoc}).
\end{tightitem}

\myparagraph{Language Detection.}
For code-level elements, we used the Google Cloud Translation API\footnote{\url{https://cloud.google.com/translate/docs/reference/rest/v3/projects/detectLanguage}} instead of \texttt{Lingua}. 
As noted in Lingua’s documentation, the model struggles with short or mixed-language inputs, a common characteristic of code comments and identifiers. 
We empirically confirmed this limitation using a manually annotated sample of 1{,}000~short texts extracted from code. 
As shown in \cref{fig:lingua_google_comparison}, Google Translator achieved higher precision and recall for detecting non-English text under the same 0.9~confidence threshold. 
We therefore adopted the Google API for code-related content, using 0.9~as the cutoff to remain consistent with our earlier setup. 
All predictions below this threshold were discarded. 
To focus on human-written content, we also removed language-specific keywords such as \texttt{int}, \texttt{float}, and \texttt{bool}, which are always English and not relevant to multilingual analysis.

\begin{figure}
\begin{center}
   \includegraphics[width=.8\linewidth, keepaspectratio]{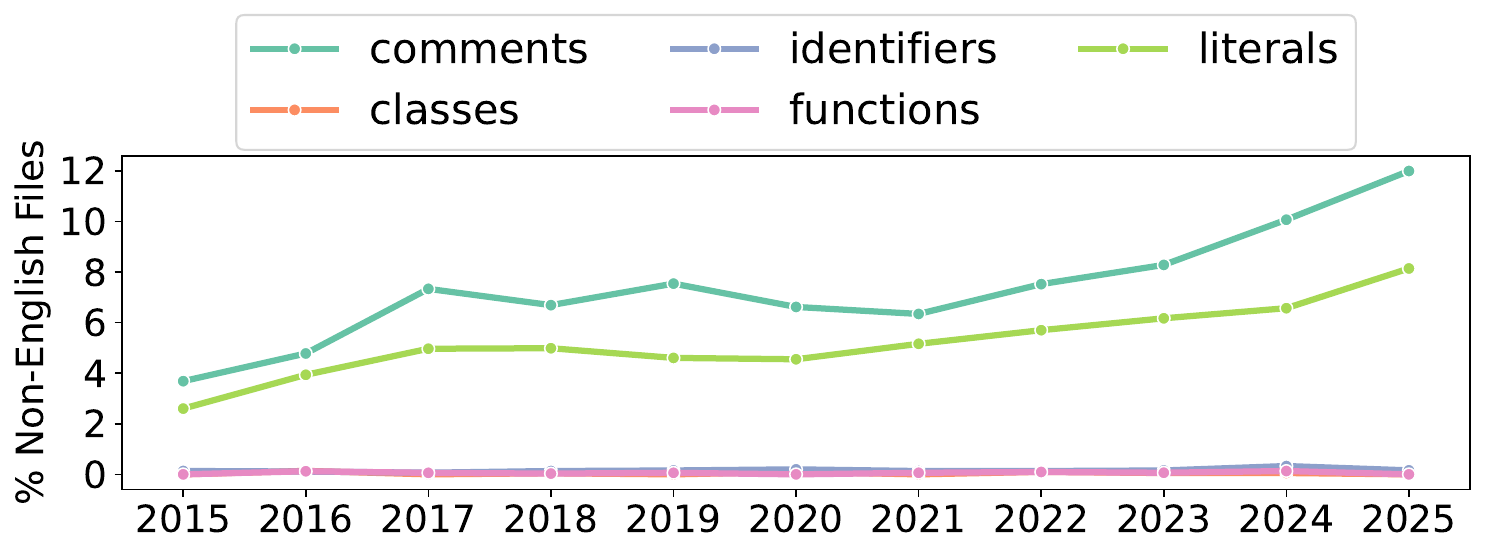}
\end{center}
\vspace{-3mm}
\caption{Share of non-English content in code elements from 2015–2025, with comments and string literals showing the most growth.}
\label{fig:code_element_rq2}
\vspace{-5mm}
\end{figure}

\myparagraph{Results.}
Our findings show a steady rise in multilingual usage across comments and string literals. These parts of the code are used for documentation or user-facing text and give developers more freedom to use their native language.
In contrast, structural elements such as identifiers, class names, and function names remain overwhelmingly English. This is likely due to two main technical constraints. First, many programming languages have restrictions on how identifiers can be named. For example, JavaScript only added support for Unicode 5.1 identifiers in ES6, which was released in 2015~\cite{exploringjsUnicode}. Unicode 5.1 itself was released in early 2008~\cite{unicodeUnicode}. Older versions of C++ also lacked full Unicode support for identifiers~\cite{cplus}. Some compilers and linters still warn against using non-ASCII characters.
Second, typing in non-Latin scripts like Arabic or Chinese often requires switching keyboard layouts or input methods. This interrupts coding and slows down development. Some tools also do not fully support multilingual code. For example, older versions of Eclipse~\cite{openjdkLoading, eclipseEclipseCommunity} and IntelliJ IDEA~\cite{jetbrainsIntelliJFile} may show broken characters when displaying mojibake text in variable names or logs, especially if encoding is not set correctly.

These limits make it harder for developers to use non-English text in core parts of code, even if they use their native language in comments or documentation.
The results reflect this pattern clearly. Comments exhibit the strongest trend, with non-English usage rising from 3.6\% of files in 2015 to 11.9\% in 2025, a 225\% relative increase. Korean comments, for instance, were present in only 0.27\% of files in 2015 but rose to 1.7\% by 2025, reflecting a steady increase in participation from Korean-speaking developers.
String literals also show a significant change, increasing from 3.2\% to 9.6\%. These elements are commonly used for documentation or user-facing text, making them more susceptible to natural language variation. One likely factor driving this recent rise is the growing use of large language models (LLMs) such as GitHub Copilot and ChatGPT. These tools often produce code with more inline comments than human-written code~\cite{park2025detection} and support multilingual input, enabling developers to write comments in their native language.

Identifiers, variables, and functions show smaller but consistent increases. The share of files with non-English identifiers grew from 0.14\% to 0.28\%, and for variables, from 0.09\% to 0.28\%. Though still marginal in absolute terms, these increases reflect a gradual broadening in naming practices. Function names remained almost entirely in English, with only negligible changes observed. Class names showed no measurable use of non-English terms throughout the period. For example, in a Japanese TypeScript project, developers used Japanese characters for both a variable and a class name in a VS Code extension plugin\footnote{\href{https://github.com/sazae657/VS-Shimonizer/blob/a71560fe03f5359a252d07db0076470dbe5cd130/src/extension.ts\#L40}{https://github.com/sazae657/VS-Shimonizer}}. In a Java testing project, one function is defined using Korean characters\footnote{\href{https://github.com/arahansa/CodeCoast30Min/blob/master/src/test/java/com/example/plaintest/EqualsTest.java}{https://github.com/arahansa/CodeCoast30Min}}. Another Python script includes a Chinese function name for a math operation\footnote{\href{https://github.com/DavidLXu/MathY/blob/b441c196cfc64e948a17fe598762cc4719f96b65/legacy/mathy-19.10.10.py}{https://github.com/DavidLXu/MathY}}. These examples are rare and do not yet reflect a large-scale shift, but they illustrate that non-English naming is both technically feasible and sometimes adopted.

\begin{figure}
\begin{center}
  \includegraphics[width=.8\linewidth, keepaspectratio]{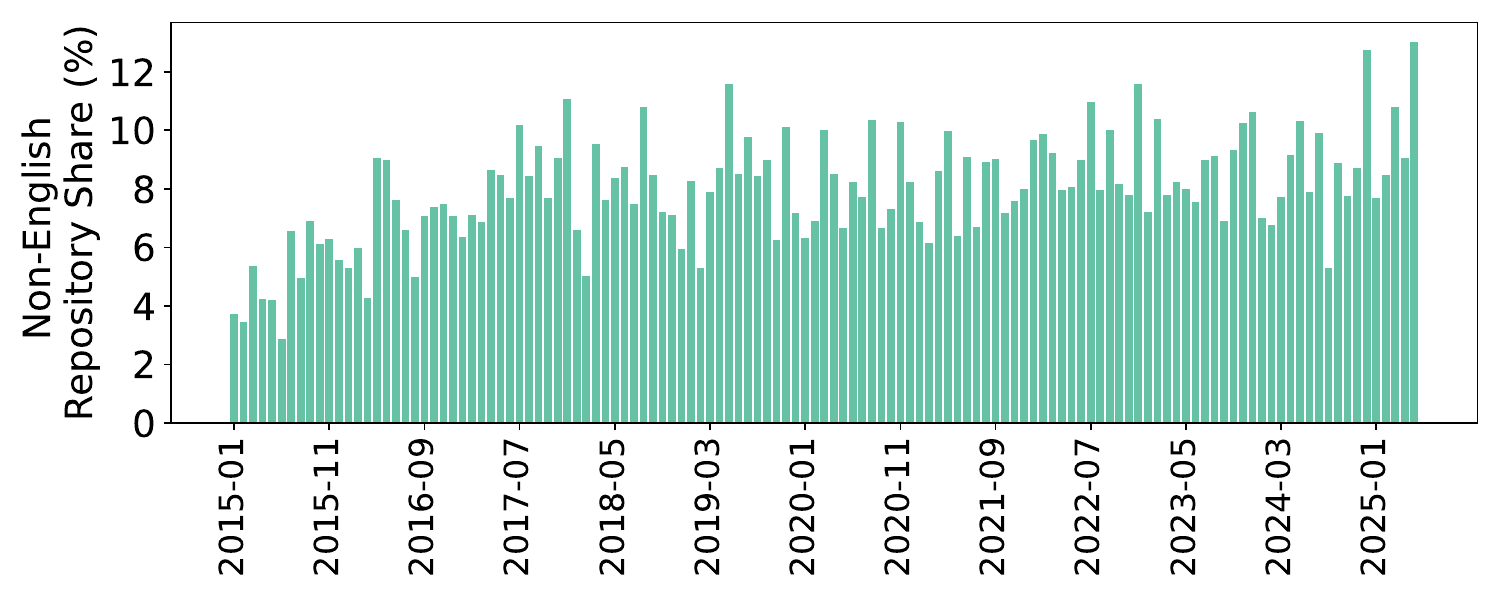}
\end{center}
\vspace{-3mm}
\caption{Percentage of repositories containing non-English documentation from 2015 to 2025.}
\label{fig:rq2_doc}
\vspace{-5mm}
\end{figure}

In addition to code-level elements, we analyzed documentation files, i.e., READMEs, to assess multilingual trends in project-level communication. Based on monthly samples from 2015 to 2025, the proportion of repositories with non-English documentation increased from 3.7\% in January 2015 to 13.0\% in May 2025, as shown in Figure~\ref{fig:rq2_doc}. The overall average across the dataset is 8.0\%. Among non-English documentation, Chinese is the most prevalent, accounting for 3.3\% of all repositories. Vietnamese documentation shows one of the most consistent upward trends, rising from 1.06\% in 2015 to 3.20\% in 2025. Russian (1.3\%), Korean (0.7\%), and Japanese (0.5\%) are also commonly observed, while Ukrainian, Thai, Greek, and Arabic collectively account for less than 0.2\%. These findings reflect a substantial shift toward multilingual documentation, reinforcing the broader trend of increasing language diversity in open-source projects. This shift can improve inclusivity by making projects more accessible to local contributors. For instance, Vietnamese README files help first-time contributors from Vietnam engage without needing English fluency. 

These results suggest that the increase in multilingualism is concentrated in human-facing components of code, particularly comments and literals. In contrast, structural identifiers such as class and function names remain almost exclusively English, indicating a persistent linguistic norm in core elements of program logic and syntax.

\section{RQ4: Does multilingualism vary across programming languages?}
\label{sec:rq4}

This research question investigates whether multilingual practices differ across programming languages and which ecosystems exhibit the most non-English content in source code and documentation.
\begin{figure}
\centering
\begin{subfigure}{0.49\linewidth}
  \includegraphics[width=\linewidth, keepaspectratio]{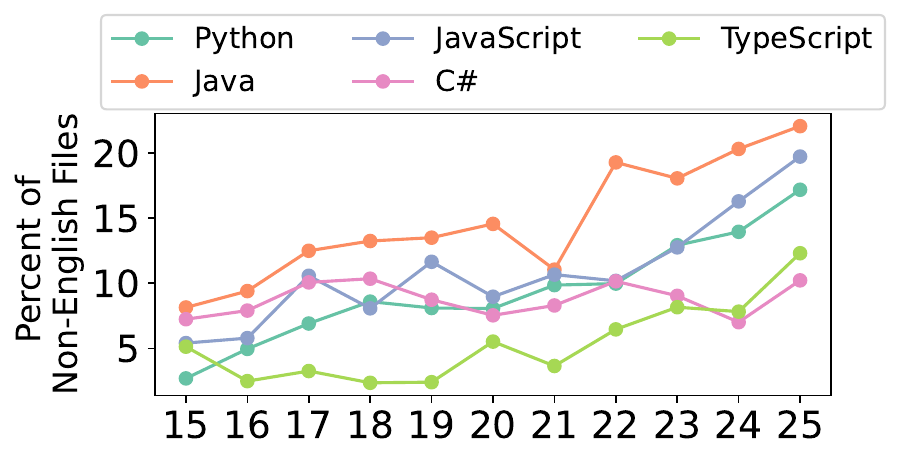}
  \vspace{-5mm}
  \caption{}
  \label{fig:fig6a_comments}
\end{subfigure}
\hfill
\begin{subfigure}{0.49\linewidth}
  \includegraphics[width=\linewidth, keepaspectratio]{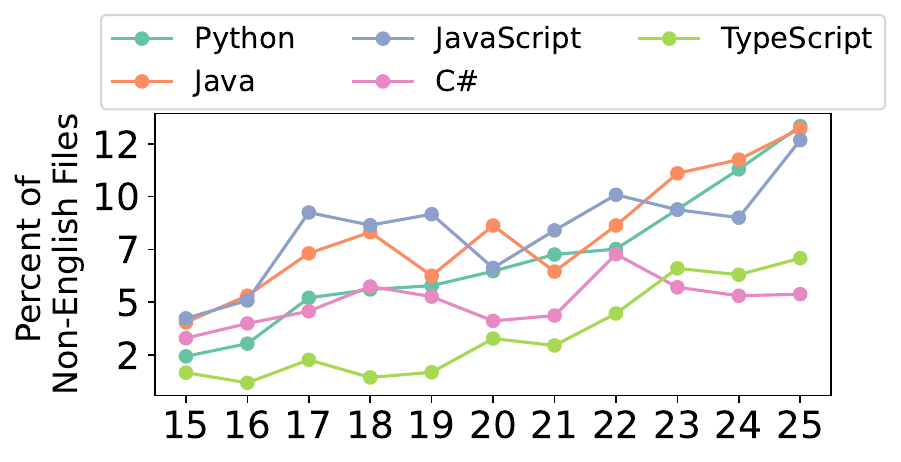}
  \vspace{-5mm}
  \caption{}
  \label{fig:fig6b_literals}
\end{subfigure}
\vspace{-5mm}
\caption{Non-English content in source code by programming language from 2015 to 2025. (a) Percentage of files with non-English comments. (b) Percentage of files with non-English string literals. Java, JavaScript, and Python exhibit the most significant multilingual growth over time.}
\label{fig:fig6}
\vspace{-5mm}
\end{figure}

\myparagraph{Data Sampling:}
For this analysis, we examined the inclusion of natural language across programming languages using the same code parsing and language detection approach described in \cref{sec:rq3}. 
To control for language-specific effects, we analyzed the same five major programming languages introduced earlier: JavaScript, Python, Java, TypeScript, and C\#. 
Because random sampling in GitHub data tends to overrepresent JavaScript and Python, we applied stratified sampling to maintain balanced representation.
Specifically, for each month from January~2015 to May~2025, we randomly selected 100~repositories per programming language, resulting in 500~repositories per month and a total of 62{,}500~repositories over the 125-month period, applying the same exclusion criteria used in~\textbf{RQ3}.

\myparagraph{Results:}
\cref{fig:fig6a_comments} and \cref{fig:fig6b_literals} show how non-English text in string literals and comments changed from 2015 to 2025. The trends differ across languages. Java has the most steady rise in multilingual content in all parts of the code that hold natural language. In 2025, 22.03\% of Java files have non-English comments and 13.2\% have non-English literals. In 2015, these numbers were 8.1\% and 4.04\%. Java’s growth likely comes from its long role in global education and enterprise work. It is still used worldwide, including in many regions where English is not the main language~\cite{nixWhereHire, jetbrainsJavaProgramming}. Java also stays common in mobile and GUI software, such as Android apps, where developers often put localized strings directly in code~\cite{androidLocalizeYour}. These use cases and the strong developer communities outside English-speaking countries help explain Java’s high multilingual share.

Python also shows steady growth in multilingual content across most elements. In 2025, 17.1\% of Python files contain non-English comments, and 13.3\% include non-English string literals. This rise matches Python’s wide global adoption, especially in education and data science~\cite{acmPythonMost, alsaggaf2022inclusion, murphy2016analysis}. Python is now the most used language worldwide, including in places where English is not the first language~\cite{githubOctoverseLeads2024}. Many schools teach Python as the first programming language, so students often write comments or literals in their own language. This learning environment, along with Python’s simplicity, likely increases its share of non-English text.

JavaScript shows a pattern similar to Java. Non-English string literals rise from 4.2\% in 2015 to 12.7\% in 2025. Comments grow from 5.4\% to 19.7\%. C\# and TypeScript show the lowest multilingual levels. In 2025, only 10.2\% of C\# files have non-English comments. TypeScript shows 12.3\% for comments, even though its non-English literals grow from 1.6\% to 7.08\%. These languages often have more stable or corporate-focused communities and slower adoption. C\# use remains tied to Microsoft and enterprise settings. For example, the official C\# Dev Kit for Visual Studio Code requires a paid license for teams with more than five developers~\cite{csDevKitFAQ}, which may limit use among small groups. TypeScript is still mostly developed and maintained by Microsoft~\cite{wikipediaTypeScriptWikipedia}, which may slow broader adoption.

Multilingual use in identifiers, class names, and function names stays rare in all languages. These elements almost never exceed 0.3\%, which aligns with common naming practices and tools that default to English~\cite{buzatu2025data}. We also see rising multilingual use in Python, JavaScript, and Java after 2022, which matches the spread of large language models like GitHub Copilot~\cite{githubGitHubCopilot} and ChatGPT~\cite{chatgptChatGPT}. Around the same time, GitHub noted that most new contributors are from non-English-speaking regions~\cite{githubOctoverseLeads2024}. LLMs can understand and produce many languages~\cite{katzy2025qualitativeinvestigationllmgeneratedmultilingual}, so they make it easier for developers to work in their native language. This likely supports the rise in multilingual content. Recent studies also show that LLMs tend to prefer Python, choosing it in 90--97\% of benchmark tasks and 58\% of project setups~\cite{twist2025llmslovepythonstudy}. This preference likely strengthens Python’s multilingual growth after 2022.

\section{RQ5: Do multilingual repositories show signs of language friction or coordination issues?}
\label{sec:rq5}

\begin{figure}
    \centering
    \includegraphics[width=.8\linewidth]{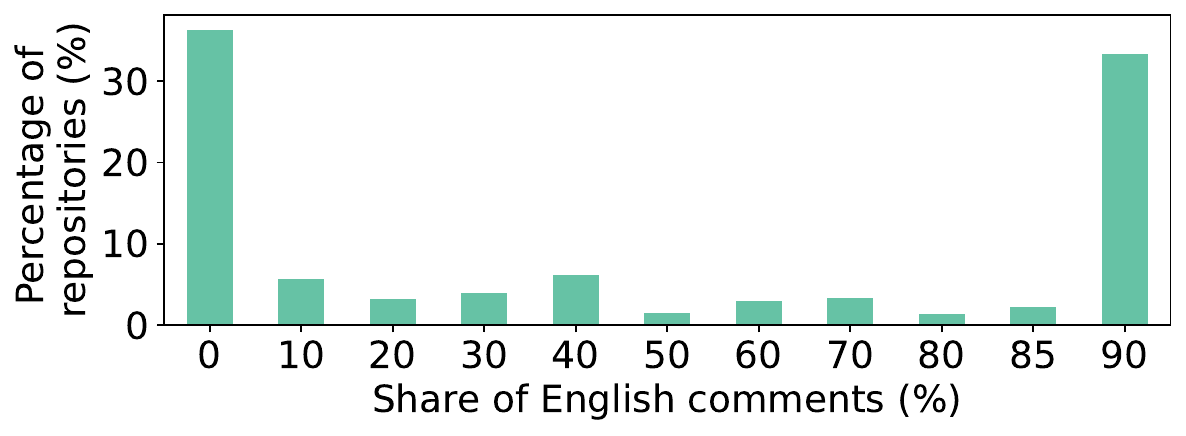}
  \vspace{-3mm}
  \caption{Distribution of English content across GitHub repositories in our dataset.}
  \label{fig:english_distribution}
  \vspace{-5mm}
\end{figure}

While RQ1 examined broad trends in the growth of non-English content across GitHub discussions, this question shifts the focus to the repository level. 
By aggregating language usage by repository, we examine whether projects operate primarily in a single dominant language or support communication in multiple languages. 
This allows us to assess whether open-source development is becoming more linguistically inclusive at the project level and whether multilingual collaboration introduces coordination challenges or fosters inclusive behavior. 

\myparagraph{Repository Classification:}
We use the same dataset as in \cref{sec:rq1}, grouping messages by repository ID and aggregating the language of comments in issues, pull requests, reviews, and discussions. 
Language detection is performed using \texttt{Lingua}, as in RQ1. 
Each repository is then classified based on the proportion of messages in each language: 
\textbf{English-dominant} if at least 90\% of all comments and issues are in English, 
\textbf{non-English-dominant} if at least 90\% are in a single non-English language, 
and \textbf{mixed-language} otherwise. 
Setting a higher threshold can potentially inflate the number of repositories labeled as multilingual; however, in our data, the distribution of English content is skewed toward the two extremes, as shown in~\cref{fig:english_distribution}. As a result, choosing a lower threshold such as 80\% or 70\% would produce nearly similar groupings, as the difference is negligible.

\begin{figure}
\begin{center}
  \includegraphics[width=.9\linewidth, keepaspectratio]{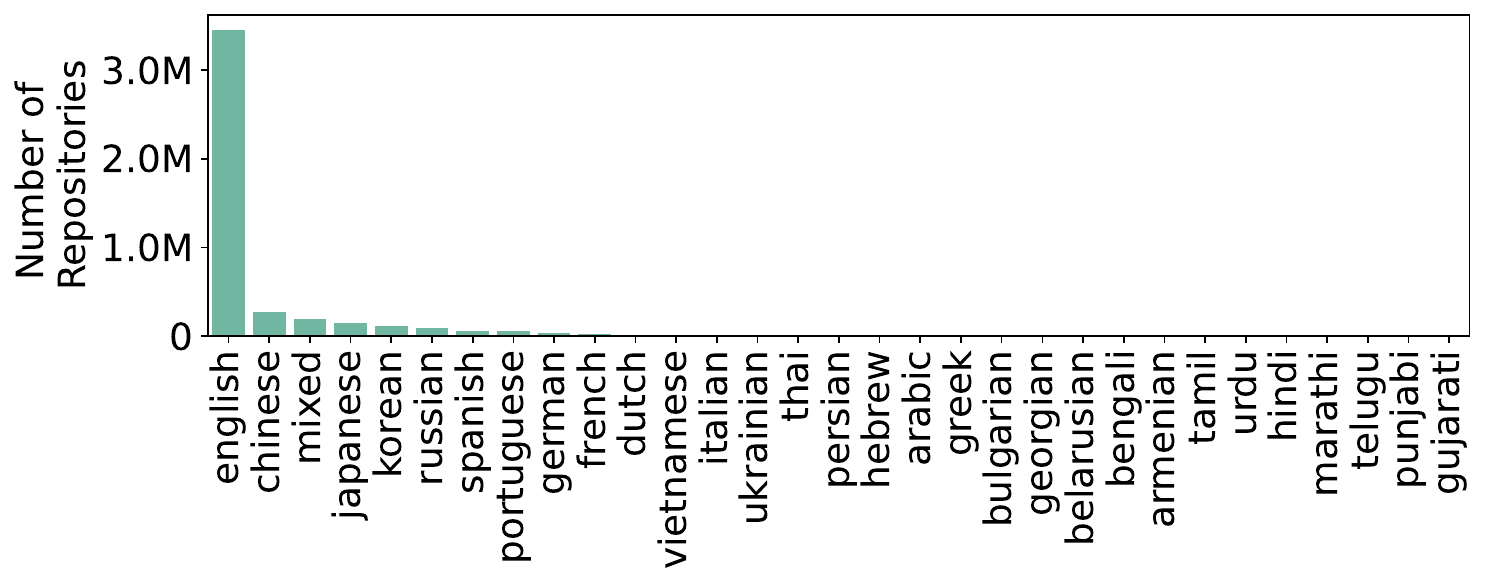}
\end{center}
\vspace{-3mm}
\caption{Number of repositories by dominant discussion language. Most multilingual activity is concentrated in English-dominant and mixed-language projects, with significant representation from Chinese, Japanese, and Korean repositories.}
\label{fig:repo_classification_rq5}
\vspace{-5mm}
\end{figure}

\myparagraph{Findings.}
As shown in~\cref{fig:repo_classification_rq5}, the majority of repositories fall into the English-dominant category, reflecting the continued central role that English plays in global open-source communication. However, we also observe a substantial number of non-English-dominant repositories, particularly in Chinese, Japanese, and Korean. These projects suggest the presence of regionally focused development communities where contributors primarily interact in their native languages.

To understand coding practices in non-English repositories, we examined 30 randomly selected Chinese and Russian projects. In the case of Chinese repositories, we found that 70\% included documentation written in Chinese. The use of Chinese within source code was somewhat lower: 40\% of the projects contained Chinese comments, and 60\% used Chinese in code elements, particularly in string literals. Similarly, among Russian repositories, 68\% had Russian documentation. Russian comments showed up in 40\%, and 28\% used Russian in code elements, especially in literals. These observations suggest that even in non-English-dominant projects, developers often retain English conventions in code while localizing documentation, comments, and interface-level strings. This finding aligns with our earlier results in~\cref{sec:rq3}, where comments and literals showed the highest multilingual growth among code elements.

While these projects demonstrate a degree of linguistic flexibility, such diversity can also introduce both opportunities and frictions. In repositories where a single language dominates, whether English or a non-English language, that language can shape participation dynamics. It sets expectations for communication, influences who feels comfortable contributing, and can either include or exclude developers based on their language proficiency. 
To understand how language is negotiated in practice, we searched for signs of language enforcement in our dataset by identifying phrases such as ``write in English,'' 
\foreignlanguage{russian}{``напишите на русском''} (write in Russian), 
\begin{CJK}{UTF8}{gbsn}
``请写中文''
\end{CJK} (write in Chinese), 
\begin{CJK}{UTF8}{min}
``日本語で書いてください''
\end{CJK} (write in Japanese), and 
\begin{CJK}{UTF8}{mj}
``한국어로 써주세요''
\end{CJK} (write in Korean).
We found 3801, 182, 1334, 717, and 149 occurrences of these phrases, respectively. 
To understand the context behind such interactions, we conducted a manual inspection of a sample of 500 issues containing these phrases.
Our analysis found that language-related comments appeared in both English-dominant and non-English-dominant repositories, though the tone and intent varied. In English-dominant repositories, such remarks were often framed as reminders or expectations to communicate in English, especially in response to issues written in other languages. In non-English-dominant repositories, similar phrases appeared more sporadically and were often expressed as suggestions aimed at improving clarity or maintaining consistency. We also observed several cases where contributors in non-English-dominant projects explicitly enforce the use of their native language, such as Russian or Chinese. These findings suggest that language enforcement is not exclusive to English speakers and can occur in any community where shared linguistic norms guide collaboration.~\cref{tab:tension_examples} lists ten representative examples of such linguistic tension, along with their corresponding GitHub issue URLs.

We also observed several cases where contributors and maintainers navigated multilingual discussions fluidly, often switching languages mid-thread to accommodate each other's preferences or limitations. For instance, in the \texttt{Buildroot} project,\footnote{\url{https://github.com/furkantokac/buildroot/issues/10\#issuecomment-920325626}} a contributor used Russian, and the maintainer explicitly welcomed it, stating they would rely on translation tools. The conversation continued in Russian, prioritizing clarity over formality. 
In another case, \texttt{mod\_execdir},\footnote{\url{https://github.com/OOPS-ORG-PHP/mod_execdir/issues/18\#issuecomment-333383366}} the issue began in English, but the maintainer noted they were more comfortable in Korean and encouraged the contributor to switch, which they did. 
Similarly, in \texttt{Far-NetBox},\footnote{\url{https://github.com/michaellukashov/Far-NetBox/issues/197}} a thread that started in English naturally transitioned into Russian, likely because both parties shared the language. 
Finally, in \texttt{MPLUS\_FONTS},\footnote{\url{https://github.com/coz-m/MPLUS_FONTS/issues/10\#issuecomment-840220932}} the maintainer encouraged the contributor to write in Japanese due to language difficulty. This led to a smooth continuation of the discussion in a bilingual Japanese–English setting.




\begin{table*}
\centering
\rowcolors{2}{gray!10}{white}
\resizebox{\textwidth}{!}{
\begin{tabular}{c|c}
\toprule
\textbf{Comment (Paraphrased)} & \textbf{GitHub URL} \\
\midrule
\begin{CJK}{UTF8}{gbsn}用中文是 Xray 社区的传统，以区分于其它喜欢用英文的项目，有些还用出优越感了\end{CJK} & \url{https://github.com/XTLS/Xray-core/issues/4348\#issuecomment-2634739760} \\

@BekzodUzb All good, but we don't understand Turkish and even machine translate doesn't help. Write in English or Russian please. & \url{https://github.com/FWGS/xash3d-fwgs/issues/1944\#issuecomment-2569944991} \\

Please write in English on our GitHub. Otherwise nobody really can read and understand what you are saying. & \url{https://github.com/flybywiresim/simbridge/issues/137\#issuecomment-2578127993} \\

Folks, writing in French here is not polite. We communicate in English. & \url{https://github.com/DurgNomis-drol/ha_toyota/issues/319\#issuecomment-2578233882} \\

\begin{CJK}{UTF8}{gbsn}这里再用中文写一遍（可能需要先生你去翻译一下）\end{CJK} & \url{https://github.com/ozntel/file-tree-alternative/issues/250\#issuecomment-2599870921} \\

\foreignlanguage{russian}{а еще в этом трекере можно писать по-русски)} & \url{https://github.com/Sectoid/nemerle/issues/1\#issuecomment-4574175} \\
\begin{CJK}{UTF8}{gbsn}... 习惯先写中文了，毕竟我是����人\end{CJK} & \url{https://github.com/X1a0He/Adobe-Downloader/issues/89\#issuecomment-2782004919}\\

\begin{CJK}{UTF8}{gbsn}建议如果写英文可以先用中文描述好后机翻一下。或者还是在 github 里直接写中文吧~~~。\end{CJK} & \url{https://github.com/RT-Thread/rt-thread/pull/10166\#issuecomment-2785061178}\\

\begin{CJK}{UTF8}{mj} PR 제목을 한국어 사용자가 읽기 편하도록 한국어로 작성하시면 어떨까요?\end{CJK} & \url{https://github.com/DaleStudy/daleui/pull/170\#issuecomment-2815349228}\\

\begin{CJK}{UTF8}{mj} 이 PR은 한국어로 작성되었으니 리뷰도 한국어로 부탁드립니다. \end{CJK} & \url{https://github.com/sparta-java-team-22/tweaty/pull/49\#issuecomment-2831811605}\\

\bottomrule
\end{tabular}
}
\caption{Examples of language-related comments and enforcement in GitHub issues.}
\label{tab:tension_examples}
\vspace{-5mm}
\end{table*}
\section{RQ6: Is Multilinguality Detrimental to Repository Engagement?}
\label{sec:rq6}
While multilingualism can support inclusivity, it may also negatively influence cooperation, i.e., how repositories are discovered, contributed to, and maintained. 
To investigate this, we analyzed four key engagement metrics across repositories categorized as English-dominant, non-English-dominant, and mixed-language: total number of comments (as a proxy for participation), number of contributors (as a measure of collaboration), number of stars (as a proxy for popularity and visibility), and issue resolution time (as a proxy for responsiveness).

\myparagraph{Methodology:}
We use the same repository classification as in RQ5, grouping all issues, pull requests, reviews, and discussion comments by repository, and labeling each as English-dominant, non-English-dominant, or mixed-language.

We define a \textit{contributor} as any non-bot GitHub user who has actively participated in a project through actions such as pushing code, opening pull requests, commenting on issues or code, creating new files, or publishing a release. This broad definition includes both code contributors and those involved in discussions, reviews, or project maintenance. Bot accounts were filtered based on known bot patterns in usernames and activity metadata. We restrict our analysis to activity occurring between January 2015 and May 2025, aligning with the overall timeline of our dataset.

Each metric is visualized on a logarithmic scale and across buckets of GitHub contribution activity, comparing distributions across language categories. 
We use this bucketed approach to control for repository size, since aggregating metrics across all projects regardless of scale could mask important differences. 
Larger repositories naturally attract more comments, stars, and contributors, so comparing projects within similar activity levels provides a clearer picture of how language influences engagement.

\myparagraph{Statistical Testing.} To ensure that the observed engagement differences are not due to random variation, we conducted statistical tests across all three engagement metrics: number of comments, number of contributors, and number of stars. Because these metrics are discrete and heavily right-skewed, we used non-parametric tests that do not assume normality. Specifically, we applied the Kruskal–Wallis H-test to compare distributions across the three language categories (English-dominant, mixed-language, and non-English-dominant). When the Kruskal–Wallis test showed significant differences, we performed pairwise Mann–Whitney U tests with Bonferroni correction to identify which groups differed. We report the corresponding H statistic, adjusted p-values, and effect sizes ($\eta^2$) for each metric.

\begin{figure}
\begin{center}
  \includegraphics[width=.9\linewidth, keepaspectratio]{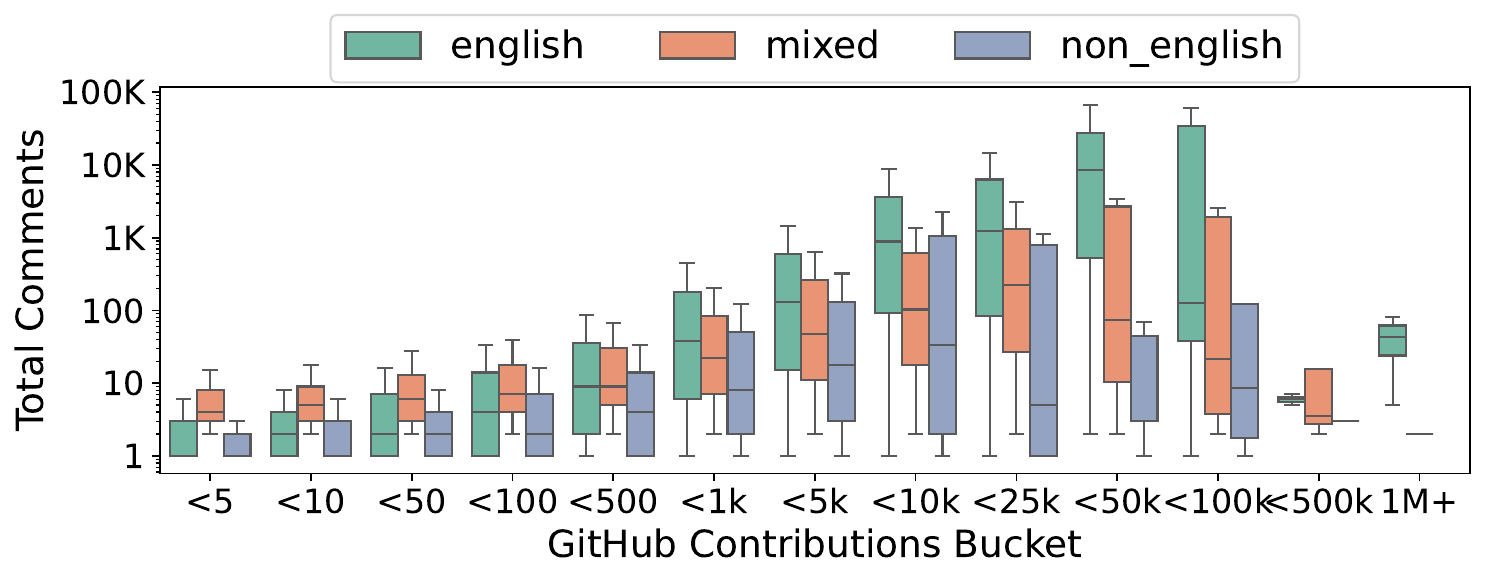}
\end{center}
\vspace{-3mm}
\caption{Total comments per repository by language category across contribution buckets (repository groups based on total commits). 
English-dominant projects lead in large buckets, while mixed-language projects show higher participation in smaller ones.}
\label{fig:comments_rq6}
\vspace{-5mm}
\end{figure}

\myparagraph{Participation: Number of Comments}
\cref{fig:comments_rq6} shows the distribution of total comments per repository across contribution buckets. In smaller repositories, mixed-language projects have a slightly higher median number of comments than English-dominant ones, suggesting more active localized discussion. However, as repository size increases, English-dominant projects surpass mixed-language repositories, both in comment volume and growth rate. The gap between English and mixed repositories widens steadily in larger buckets, with English-dominant repositories showing the highest levels of discussion beyond 1k contributions. Non-English-dominant repositories consistently show the lowest median comment counts across all size categories.
This pattern suggests that non-English repositories may experience reduced participation in collaborative discussions. Since GitHub comments are central to issue triage, bug reporting, feature discussion, and community feedback, lower comment volume may hinder transparent collaboration or discourage new contributors from engaging.

The Kruskal–Wallis test confirmed that the observed differences are statistically significant ($H = 101{,}897.31$, $p < 0.001$, $\eta^2 = 0.1859$). Pairwise Mann–Whitney U tests with Bonferroni correction further showed that English-dominant repositories receive significantly more comments than both mixed and non-English repositories ($p < 0.001$). These results indicate a large practical effect, with language category explaining a substantial share of the variation in participation.

\begin{figure}
\begin{center}
  \includegraphics[width=.9\linewidth, keepaspectratio]{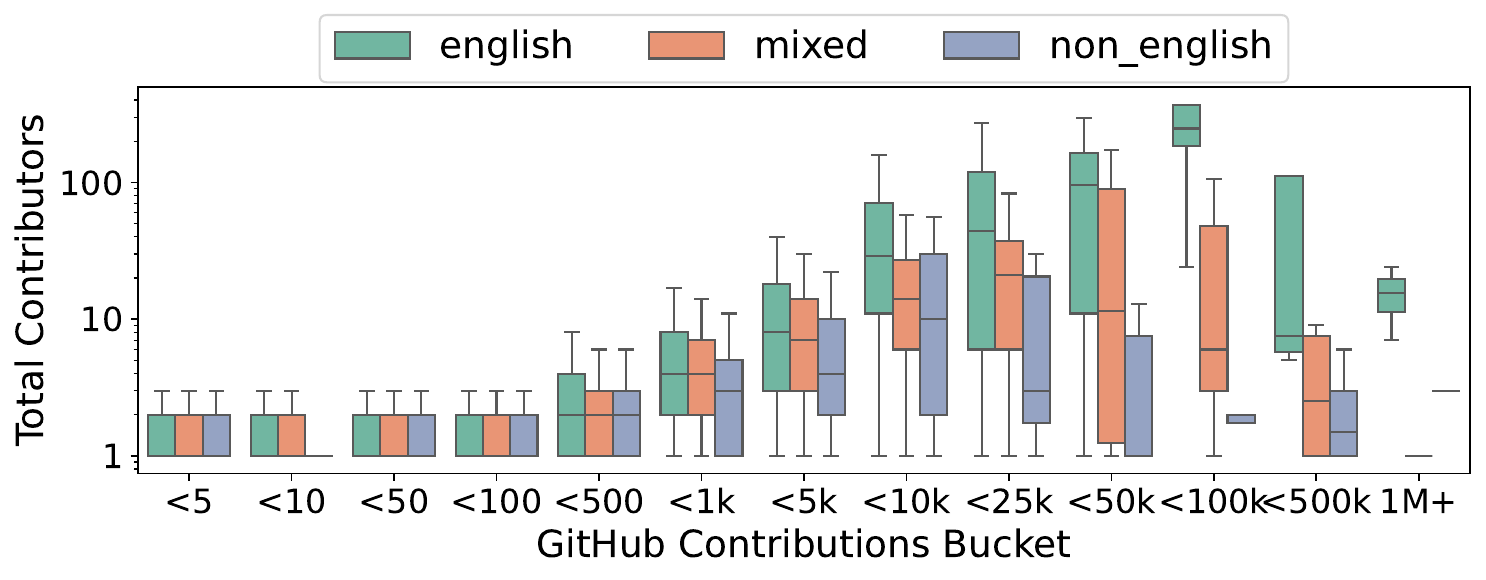}
\end{center}
\vspace{-3mm}
\caption{Distribution of unique contributors per repository across language categories. English repositories have more contributor growth as project size increases, while non-English repositories consistently have fewer contributors.}
\label{fig:contributors_rq6}
\vspace{-5mm}
\end{figure}
\myparagraph{Collaboration: Number of Contributors}
To assess collaboration, we count the number of unique contributors per repository. Figure~\ref{fig:contributors_rq6} shows the contributor distribution across language categories. In smaller repositories, contributor counts are generally low and similar across all groups. This is expected, as many small projects are maintained by a single developer or a very small team~\cite{redmonkSizeOpensource, gnuWhenFree}. Contributions often come from a few users who report bugs, request features, or submit patches. As a result, the median contributor count stays low regardless of the language used.
As repositories grow, differences between language groups become clearer. English-dominant projects show steady contributor growth, likely due to greater visibility and broader accessibility. They may attract a more geographically diverse contributor base. Mixed-language repositories show slower growth, even with multilingual participation. This may result from communication challenges (\cref{sec:rq5}), inconsistent norms, or unclear language policies that discourage contributors.
While causality is hard to confirm, factors like popularity, responsiveness, repository age, and community size may influence these patterns. Further study is needed to examine these.
Non-English-dominant repositories consistently show the fewest contributors across all sizes. These projects often serve regional or linguistic communities and may be less visible on GitHub. While this focused scope doesn’t imply lower quality, it can limit collaborative growth. For example, the French project \texttt{ratpstatus}~\cite{laurentRATPStatus} tracks transport disruptions in Île-de-France and is active, yet has only two contributors. The Chinese \texttt{API} project~\cite{yuncaijiAPI} offers reverse data interfaces with frequent commits, also by just two contributors. These examples show how language barriers and low discoverability can discourage external contributors. Using a local language can help serve a specific community, but comes with the cost of reduced visibility. Developers should consider this tradeoff. Language choice affects who participates. Prior work shows linguistic divides hinder global collaboration~\cite{wang2024uncovering}, while inclusive communication helps attract and retain diverse contributors~\cite{hyrynsalmi2024bridging}.

The Kruskal–Wallis test confirmed a statistically significant difference in contributor counts across language categories ($H = 6{,}509.79$, $p < 0.001$, $\eta^2 = 0.0194$). Pairwise Mann–Whitney U tests with Bonferroni correction showed that English-dominant repositories have significantly more contributors than both mixed and non-English repositories ($p < 0.001$). Although the effect size is smaller than for participation, the difference remains systematic, indicating that language orientation influences collaboration at scale.

\begin{figure}
\begin{center}
  \includegraphics[width=.9\linewidth, keepaspectratio]{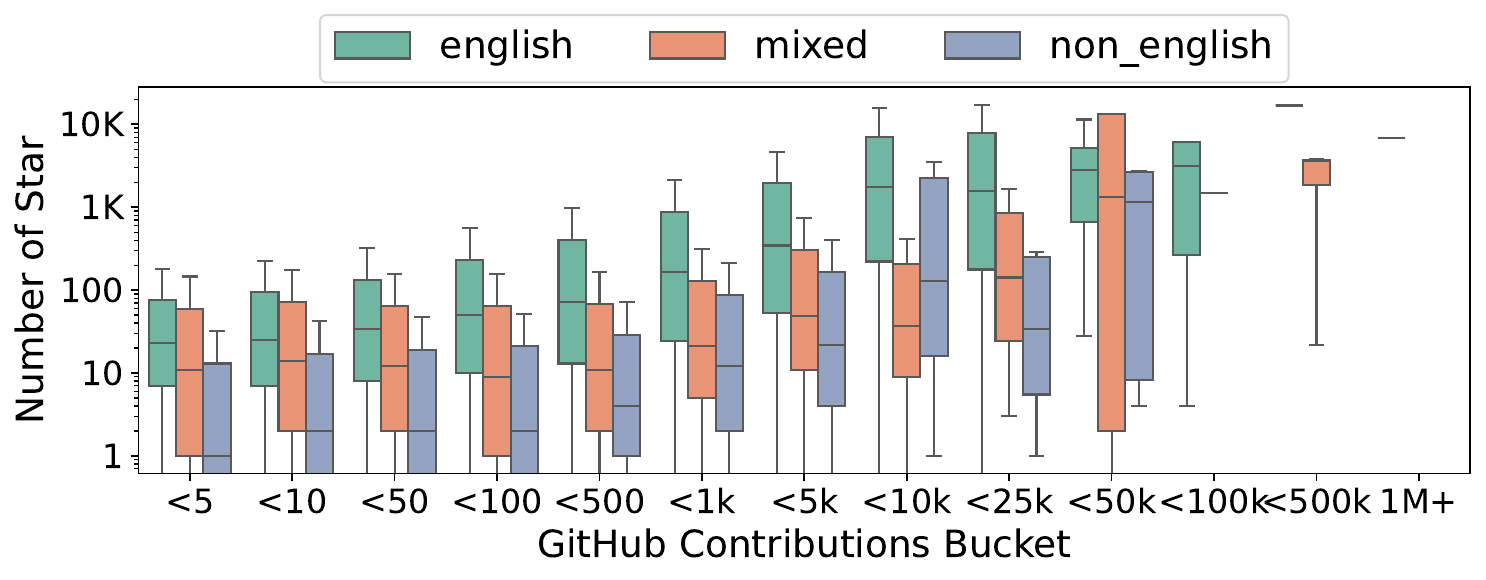}
\end{center}
\vspace{-3mm}
\caption{Distribution of GitHub stars per repository across language categories and contribution levels. English-dominant repositories show higher visibility, especially as project size increases.}
\label{fig:stars_rq6}
\vspace{-5mm}
\end{figure}

\myparagraph{Visibility: Number of Stars}
GitHub stars are often used as a proxy for a project's popularity and reach. Prior work shows that stars signal user interest and are widely used in empirical studies to rank and select repositories~\cite{borges2016understanding, ray2014large, borges2018s, koch2024fault, borges2016predicting}. Stars also have practical value: users bookmark projects they find useful~\cite{zhang2017detecting}, and maintainers highlight star counts to attract contributors or funders~\cite{dabbish2012social}. A survey by Borges et al.~\cite{borges2018s} reports that most developers look at stars before using or contributing to a project.
\cref{fig:stars_rq6} shows the distribution of GitHub stars across repositories grouped by language category and contribution size. In small repositories (fewer than 100 contributions), differences in star counts are modest. But as activity increases, English-dominant repositories attract substantially more stars, dominating the upper visibility tiers. Mixed-language repositories follow, showing moderate growth, while non-English-dominant repositories consistently receive the fewest stars across all sizes. This highlights a clear visibility gap tied to language, with English projects more likely to gain recognition on the platform.

Unlike contributor count, which reflects actual participation, stars reflect perceived value or utility and are shaped by visibility. Mixed-language repositories gain more traction than non-English ones but still fall behind English projects. Non-English repositories remain concentrated in lower star ranges. Projects that do not use English may struggle to attract attention even when active and well-maintained. Because highly starred repositories are more likely to be recommended or referenced, this gap reinforces platform inequality.


The Kruskal–Wallis test confirmed a significant difference in star counts across language categories ($H = 9{,}279.26$, $p < 0.001$, $\eta^2 = 0.1441$). Pairwise Mann–Whitney U tests with Bonferroni correction showed that English-dominant repositories have significantly more stars than mixed and non-English repositories, and mixed repositories more than non-English ones ($p < 0.001$; medians: $36$, $12$, and $2$, respectively). The large effect size indicates that language category strongly influences repository visibility and popularity.

\begin{figure}
\begin{center}
  \includegraphics[width=.9\linewidth, keepaspectratio]{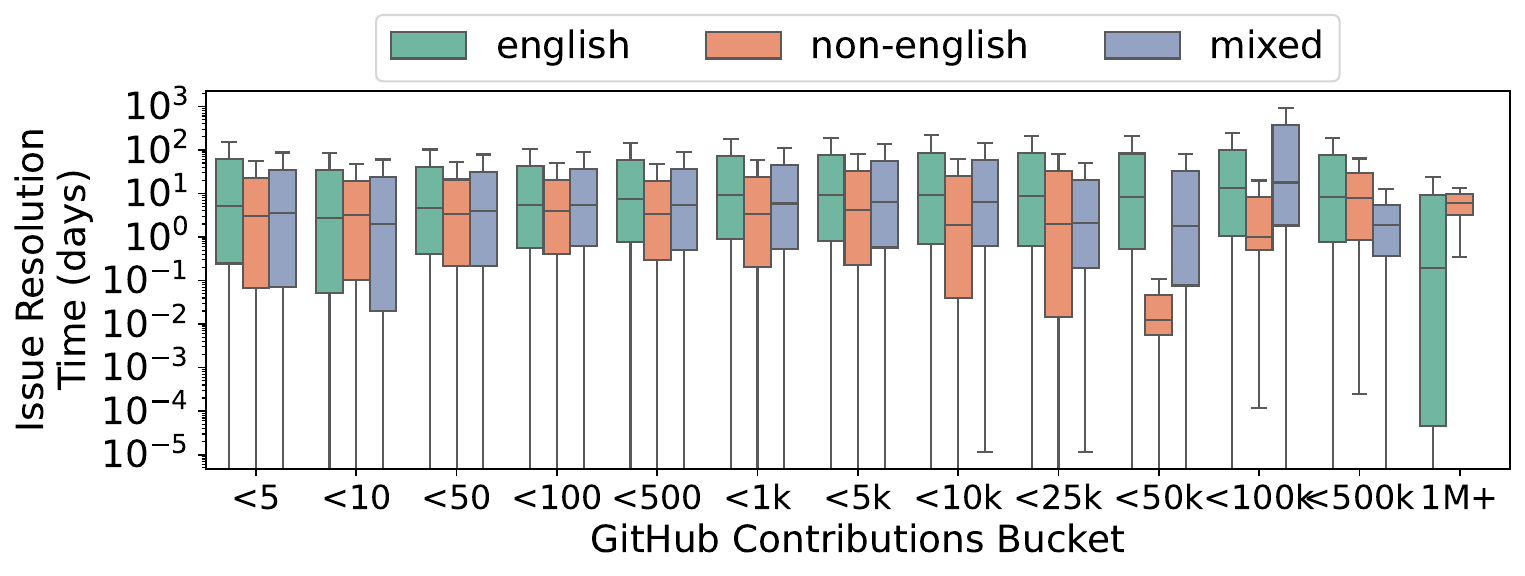}
\end{center}
\vspace{-3mm}
\caption{Distribution of issue resolution time across language categories.}
\label{fig:issue_resolution_rq6}
\vspace{-5mm}
\end{figure}

\myparagraph{Responsiveness: Issue Resolution Time}
Issue resolution time is a key indicator of project responsiveness and maintenance quality. It reflects how quickly developers react to user feedback, bug reports, or feature requests~\cite{zhang2013predicting, gousios2014exploratory}. Previous studies show that long delays can slow progress and reduce overall project performance~\cite{hasan2023understandingtimeresponsegithub}. Figure~\ref{fig:issue_resolution_rq6} shows the distribution of issue resolution times across language categories. English-dominant repositories take the longest to resolve issues, followed by mixed-language, and then non-English repositories. A Kruskal–Wallis test confirmed that these differences are statistically significant ($H = 417{,}873.16$, $p < 0.001$, $\eta^2 = 0.0065$). Pairwise Mann–Whitney U tests show significant differences across all pairs ($p < 0.001$), with median times of 7.17~days for English projects, 4.70~days for mixed projects, and 3.39~days for non-English projects.

While the effect size is small, the trend is systematic and somewhat counterintuitive: one might expect English-dominant projects, with larger contributor pools, to resolve issues faster. Instead, they often receive more issues and face coordination overhead due to scale and global participation~\cite{vasilescu2015gender, wachs2022geography}. Non-English projects are often smaller and community-focused, which can lead to quicker responses within narrower scopes. Another possible explanation is that some countries, such as Korea, actively support or fund OSS in local languages, which could influence responsiveness. This remains a hypothesis, and future work should examine it further. These findings suggest that linguistic orientation not only shapes collaboration but also affects how efficiently projects address reported problems.

\section{Threats to Validity}
\myparagraph{Language Detection Accuracy: }
Detecting the natural language of short, informal GitHub text is challenging, especially for code-mixed messages or short comments~\cite{bergsma2012language, barman2014code, kostic2023monolingual}. Models may struggle with short or noisy inputs. To mitigate errors, we use Lingua and the Google Translate API in strict confidence mode, accepting predictions only when confidence exceeds 0.9. We restrict analysis to non-Latin languages because Latin-based languages (e.g., English, Spanish, French) share vocabulary that often confuses detectors on short text. Our estimates are therefore conservative lower bounds.

\myparagraph{Filtering Criteria.}
For code-level analysis, we included only patches that introduced at least one new file and contained more than 500 characters. 
As our goal is to study longitudinal change, separating small edits or mixed-language content within the same file is difficult for reliable code parsing, which further motivates focusing on newly added files. 
Although this filter may exclude some meaningful edits, such as documentation updates or refactorings, applying it uniformly across all time periods and languages preserves comparability.

\myparagraph{Partial Contribution History.}
All participation metrics, such as the number of comments, stars, or contributors, are calculated based on activity observed between January 2015 and May 2025. Any comments, stars, or contributions made before 2015 or after May 2025 are not considered. This temporal constraint may undercount activity, particularly for older repositories or projects with long histories. However, we apply the same time window across all language groups, so we believe that comparisons remain valid within the scope of the dataset. Additionally, we believe this dataset remains representative due to the statistically significant sample size of contributions collected over the decade-long period, which captures a substantial portion of activity across diverse repositories.


\myparagraph{Repository Sampling and Classification:}
For code-level multilingual analysis (RQ3, RQ4, RQ5), we sample repositories and code patches from monthly GitHub activity. This ensures temporal and language balance but may miss less active or less visible repositories. Additionally, we classify repositories as English-dominant, non-English-dominant, or mixed based on issue and comment data, using a 90\% threshold for language dominance. In practice, this threshold has minimal impact on group composition because the distribution of English content is strongly skewed toward the extremes, meaning lower thresholds (e.g., 80\% or 70\%) would produce similar groupings.


\myparagraph{Generalizability:}
Our findings are based on GitHub and may not generalize to platforms such as GitLab or Bitbucket, or to independently hosted open-source projects. GitHub usage also varies by region and has faced access restrictions or bans in countries such as China, India, and Russia~\cite{githubCensorshipWiki}, as well as sanctions-related restrictions affecting regions including Iran, Syria, and Crimea~\cite{githubBanSanctioned2019}. These factors may limit visibility and participation. Finally, our analysis covers only public repositories; behavior in private repositories may differ.
\section{Discussion}
The rise of multilinguality in open-source repositories marks a shift in global software development. While it increases accessibility for non-English-speaking developers, it also introduces challenges for maintainers, researchers, and developers who must adapt to support broader participation.

\myparagraph{Beyond English in Developer Research:}
Much of the current research on developer communication, whether focused on toxicity~\cite{miller2022did, ToxiSpanSE, landscapetoxic}, sexism~\cite{sultana2024assessinginfluencetoxicgender}, politeness~\cite{destefanis2018measuring}, or emotional tone~\cite{murgia2018exploratory}, has been conducted exclusively on English-language data. This creates a blind spot in the community's understanding of global developer behavior. As our results show, a significant portion of open source interaction now happens in other languages. Tools and models developed on English-only corpora risk missing or misclassifying key signals in multilingual settings.
Future work should explore language-specific forms of toxicity, inclusion, and engagement. This may require building labeled datasets in non-English languages, adapting sentiment or toxicity classifiers for regional communication norms, and evaluating how moderation practices vary across linguistic communities.

\myparagraph{Multilingual Tooling and Infrastructure:}
Automation is central to modern workflows, from issue triage to documentation bots~\cite{wessel2018power, santhanam2022bots, golzadeh2020bot, golzadeh2021ground, golzadeh2021identifying}. Many tools assume English input, limiting their usefulness in multilingual repositories. New tooling should support translation-aware linting, multilingual bots, localized search, and documentation generation. IDEs could offer localized interfaces and clearer error messages. Compilers and programming languages could provide better diagnostic support for multilingual users.


\myparagraph{Community Tensions Around Language Use:}
As multilingual participation grows, contributors negotiate which language to use and when. Our RQ5 findings show that language enforcement is a recurring issue in GitHub repositories. Phrases like ``write in English'' or \begin{CJK}{UTF8}{gbsn}``请写中文''\end{CJK} reveal ongoing frictions around communication norms. This enforcement is not limited to English-dominant projects; contributors in non-English repositories also assert language preferences. Multilingualism does not remove gatekeeping but shifts its focus, as language choice can include or exclude depending on how norms are set.
Whether multilingual collaboration becomes more inclusive or more fragmented depends on project culture and governance. Some maintainers use translation tools or switch languages to accommodate contributors, while others reject messages or ask users to follow a preferred language. Future work could explore strategies such as language labels, integrated translation support, or multilingual contribution guidelines to manage these tensions more effectively.

\myparagraph{The Visibility and Contribution Tradeoff: }
Despite the advantages of multilingualism, our findings show a drawback: repositories using non-English languages tend to receive fewer stars, attract fewer contributors, and generate fewer comments. This pattern is strongest in larger, active projects. English remains the lingua franca of open source, and moving away from it can reduce a project's discoverability. These metrics matter because they signal popularity and activity to funders, contributors, and users~\cite{borges2018s, dabbish2012social, he20244}. Projects outside the English-dominated space may struggle to gain traction even when serving important regional needs.
Developers managing multilingual projects should be aware of this visibility gap. When possible, English summaries, dual-language READMEs, or tagging issues for translation can help bridge linguistic divides. There is no single solution: some projects may prefer regional accessibility over global reach. What matters is making intentional language choices rather than defaulting to one approach.

\myparagraph{Demographics and Linguistic Dynamics in OSS:}
Prior work on inclusion in OSS often focuses on demographics such as ethnicity, gender, or geography~\cite{bosu2019diversity, santos2025understanding}. These factors matter, but language and economics also significantly shape participation, and they are largely overlooked in the literature. Demographic trends do not directly correspond to linguistic behavior. Countries like India and Pakistan have multiple official languages, yet English is widely used as the working lingua franca. As a result, demographic counts alone cannot portray actual inclusion patterns. These linguistic choices reflect cultural, educational, and economic forces that demographic indicators cannot capture. Future research on inclusivity should consider these nuances rather than relying solely on demographic measures.

\myparagraph{AI and the Future of Multilingual Onboarding:}
Language barriers are known to hinder onboarding in open-source communities. Prior work shows that newcomers struggle to participate when project communication is English-centric~\cite{steinmacher2015social}, and mentoring relationships often break down when contributors lack shared language proficiency~\cite{balali2018newcomers}. As multilingual participation grows, these challenges may intensify. AI tools offer a possible mitigation path: automatic translation, summarization, and code-aware rewriting could help newcomers read and write project artifacts in their preferred language, lowering the entry barrier. At the same time, AI may also amplify linguistic tensions. Developers using AI assistants may generate more non-English issues, comments, and documentation, increasing the burden on maintainers and reinforcing the language negotiations we observe in mixed-language repositories. How communities integrate AI-mediated translation will shape whether multilingual collaboration becomes smoother or more fragmented.
\section{Data Availability}

The dataset for this study comes from public GitHub activity between January 2015 and May 2025, collected via GHArchive and the GitHub REST API. Due to the data volume, we do not redistribute raw messages. We release aggregated statistics, language annotations, and repository-level classifications at: 
\begin{center}
\url{https://github.com/kal-purush/bhasha-nirikhhon}
\end{center}
The repository includes preprocessing scripts, language detection outputs, and selected metadata to support replication.



\section{Conclusion}
This paper presents the first large-scale study of multilingualism in open source software development across both communication and code. By analyzing \numberofcomments GitHub discussions and over \numberofrepos code patches from 2015 to May 2025, we show that multilingual participation is steadily increasing, especially in Korean, Chinese, and Russian. This shift appears not only in issues and comments but also in code elements such as literals, comments, and documentation.
Our results suggest that open source is gradually becoming more inclusive of non-English speakers. At the same time, the trend introduces new challenges. We observe language-based tensions around communication norms, and multilingual or non-English repositories tend to receive fewer contributions, comments, and stars. We also find a counterintuitive pattern in issue resolution: English-dominant projects resolve issues the slowest, while non-English projects are the fastest.
Taken together, these findings highlight a tradeoff between linguistic expressivity and community inclusivity on one side, and visibility, collaboration, and coordination on the other. They point to the need for more language-aware tools, analyses, and future research on how multilingualism shapes developer experience, project success, and community dynamics.


\bibliographystyle{ACM-Reference-Format}
\bibliography{sample-base}


\begin{thebibliography}{96}


\ifx \showCODEN    \undefined \def \showCODEN     #1{\unskip}     \fi
\ifx \showISBNx    \undefined \def \showISBNx     #1{\unskip}     \fi
\ifx \showISBNxiii \undefined \def \showISBNxiii  #1{\unskip}     \fi
\ifx \showISSN     \undefined \def \showISSN      #1{\unskip}     \fi
\ifx \showLCCN     \undefined \def \showLCCN      #1{\unskip}     \fi
\ifx \shownote     \undefined \def \shownote      #1{#1}          \fi
\ifx \showarticletitle \undefined \def \showarticletitle #1{#1}   \fi
\ifx \showURL      \undefined \def \showURL       {\relax}        \fi
\providecommand\bibfield[2]{#2}
\providecommand\bibinfo[2]{#2}
\providecommand\natexlab[1]{#1}
\providecommand\showeprint[2][]{arXiv:#2}

\bibitem[Alebachew et~al\mbox{.}(2025)]%
        {alebachew2025pageexaminingdeveloperperception}
\bibfield{author}{\bibinfo{person}{Yoseph~Berhanu Alebachew}, \bibinfo{person}{Minhyuk Ko}, {and} \bibinfo{person}{Chris Brown}.} \bibinfo{year}{2025}\natexlab{}.
\newblock \showarticletitle{Are We on the Same Page? Examining Developer Perception Alignment in Open Source Code Reviews}. In \bibinfo{booktitle}{\emph{Proceedings of the 29th International Conference on Evaluation and Assessment in Software Engineering}}. \bibinfo{pages}{57--67}.
\newblock


\bibitem[Alsaggaf et~al\mbox{.}(2022)]%
        {alsaggaf2022inclusion}
\bibfield{author}{\bibinfo{person}{Wafaa Alsaggaf}, \bibinfo{person}{H Baaqeel}, \bibinfo{person}{S Alkhuraiji}, {and} \bibinfo{person}{Hani Brdesee}.} \bibinfo{year}{2022}\natexlab{}.
\newblock \showarticletitle{The Inclusion of Python as Introductory Computer Programming in the Preparatory Year of Higher Education: Modeling for Students’ Perceptions}.
\newblock \bibinfo{journal}{\emph{IJCSNS International Journal of Computer Science and Network Security}} \bibinfo{volume}{22}, \bibinfo{number}{3} (\bibinfo{year}{2022}), \bibinfo{pages}{565--574}.
\newblock


\bibitem[{Android Developers}(2024)]%
        {androidLocalizeYour}
\bibfield{author}{\bibinfo{person}{{Android Developers}}.} \bibinfo{year}{2024}\natexlab{}.
\newblock \bibinfo{title}{{Localize Your App | App Architecture | Android Developers}}.
\newblock \bibinfo{howpublished}{\url{https://developer.android.com/guide/topics/resources/localization}}.
\newblock


\bibitem[Avelino et~al\mbox{.}(2016)]%
        {Avelino_2016}
\bibfield{author}{\bibinfo{person}{Guilherme Avelino}, \bibinfo{person}{Leonardo Passos}, \bibinfo{person}{Andre Hora}, {and} \bibinfo{person}{Marco~Tulio Valente}.} \bibinfo{year}{2016}\natexlab{}.
\newblock \showarticletitle{A novel approach for estimating Truck Factors}. In \bibinfo{booktitle}{\emph{2016 IEEE 24th International Conference on Program Comprehension (ICPC)}}. \bibinfo{publisher}{IEEE}, \bibinfo{pages}{1–10}.
\newblock
\href{https://doi.org/10.1109/icpc.2016.7503718}{doi:\nolinkurl{10.1109/icpc.2016.7503718}}


\bibitem[Balali et~al\mbox{.}(2018)]%
        {balali2018newcomers}
\bibfield{author}{\bibinfo{person}{Sogol Balali}, \bibinfo{person}{Igor Steinmacher}, \bibinfo{person}{Umayal Annamalai}, \bibinfo{person}{Anita Sarma}, {and} \bibinfo{person}{Marco~Aurelio Gerosa}.} \bibinfo{year}{2018}\natexlab{}.
\newblock \showarticletitle{Newcomers’ barriers... is that all? an analysis of mentors’ and newcomers’ barriers in OSS projects}.
\newblock \bibinfo{journal}{\emph{Computer Supported Cooperative Work (CSCW)}} \bibinfo{volume}{27}, \bibinfo{number}{3} (\bibinfo{year}{2018}), \bibinfo{pages}{679--714}.
\newblock


\bibitem[Barman et~al\mbox{.}(2014)]%
        {barman2014code}
\bibfield{author}{\bibinfo{person}{Utsab Barman}, \bibinfo{person}{Amitava Das}, \bibinfo{person}{Joachim Wagner}, {and} \bibinfo{person}{Jennifer Foster}.} \bibinfo{year}{2014}\natexlab{}.
\newblock \showarticletitle{Code mixing: A challenge for language identification in the language of social media}. In \bibinfo{booktitle}{\emph{Proceedings of the first workshop on computational approaches to code switching}}. \bibinfo{pages}{13--23}.
\newblock


\bibitem[Becker(2019)]%
        {becker2019parlez}
\bibfield{author}{\bibinfo{person}{Brett~A Becker}.} \bibinfo{year}{2019}\natexlab{}.
\newblock \showarticletitle{Parlez-vous Java? Bonjour La Monde!= Hello World: Barriers to Programming Language Acquisition for Non-Native English Speakers.}. In \bibinfo{booktitle}{\emph{PPIG}}.
\newblock


\bibitem[Becker et~al\mbox{.}(2018)]%
        {becker2018fix}
\bibfield{author}{\bibinfo{person}{Brett~A Becker}, \bibinfo{person}{Cormac Murray}, \bibinfo{person}{Tianyi Tao}, \bibinfo{person}{Changheng Song}, \bibinfo{person}{Robert McCartney}, {and} \bibinfo{person}{Kate Sanders}.} \bibinfo{year}{2018}\natexlab{}.
\newblock \showarticletitle{Fix the first, ignore the rest: Dealing with multiple compiler error messages}. In \bibinfo{booktitle}{\emph{Proceedings of the 49th ACM technical symposium on computer science education}}. \bibinfo{pages}{634--639}.
\newblock


\bibitem[Bergsma et~al\mbox{.}(2012)]%
        {bergsma2012language}
\bibfield{author}{\bibinfo{person}{Shane Bergsma}, \bibinfo{person}{Paul McNamee}, \bibinfo{person}{Mossaab Bagdouri}, \bibinfo{person}{Clayton Fink}, {and} \bibinfo{person}{Theresa Wilson}.} \bibinfo{year}{2012}\natexlab{}.
\newblock \showarticletitle{Language identification for creating language-specific twitter collections}. In \bibinfo{booktitle}{\emph{Proceedings of the second workshop on language in social media}}. \bibinfo{pages}{65--74}.
\newblock


\bibitem[Berkholz(2013)]%
        {redmonkSizeOpensource}
\bibfield{author}{\bibinfo{person}{Donnie Berkholz}.} \bibinfo{year}{2013}\natexlab{}.
\newblock \bibinfo{title}{{The Size of Open-Source Communities and Its Impact Upon Activity, Licensing, and Hosting --- redmonk.com}}.
\newblock \bibinfo{howpublished}{\url{https://redmonk.com/dberkholz/2013/04/22/the-size-of-open-source-communities-and-its-impact-upon-activity-licensing-and-hosting/}}.
\newblock


\bibitem[Bhuiyan et~al\mbox{.}(2025)]%
        {bhuiyan2025not}
\bibfield{author}{\bibinfo{person}{Masudul Hasan~Masud Bhuiyan}, \bibinfo{person}{Matteo Varvello}, \bibinfo{person}{Yasir Zaki}, {and} \bibinfo{person}{Cristian-Alexandru Staicu}.} \bibinfo{year}{2025}\natexlab{}.
\newblock \showarticletitle{{Not All Visitors are Bilingual: A Measurement Study of the Multilingual Web from an Accessibility Perspective}}. In \bibinfo{booktitle}{\emph{Proceedings of the 2025 ACM Internet Measurement Conference}} (USA) \emph{(\bibinfo{series}{IMC '25})}. \bibinfo{publisher}{Association for Computing Machinery}, \bibinfo{address}{New York, NY, USA}, \bibinfo{pages}{871–881}.
\newblock
\showISBNx{9798400718601}
\href{https://doi.org/10.1145/3730567.3764505}{doi:\nolinkurl{10.1145/3730567.3764505}}


\bibitem[Borges et~al\mbox{.}(2016a)]%
        {borges2016predicting}
\bibfield{author}{\bibinfo{person}{Hudson Borges}, \bibinfo{person}{Andre Hora}, {and} \bibinfo{person}{Marco~Tulio Valente}.} \bibinfo{year}{2016}\natexlab{a}.
\newblock \showarticletitle{Predicting the popularity of github repositories}. In \bibinfo{booktitle}{\emph{Proceedings of the The 12th international conference on predictive models and data analytics in software engineering}}. \bibinfo{pages}{1--10}.
\newblock


\bibitem[Borges et~al\mbox{.}(2016b)]%
        {borges2016understanding}
\bibfield{author}{\bibinfo{person}{Hudson Borges}, \bibinfo{person}{Andre Hora}, {and} \bibinfo{person}{Marco~Tulio Valente}.} \bibinfo{year}{2016}\natexlab{b}.
\newblock \showarticletitle{Understanding the factors that impact the popularity of GitHub repositories}. In \bibinfo{booktitle}{\emph{2016 IEEE international conference on software maintenance and evolution (ICSME)}}. IEEE, \bibinfo{pages}{334--344}.
\newblock


\bibitem[Borges and Valente(2018)]%
        {borges2018s}
\bibfield{author}{\bibinfo{person}{Hudson Borges} {and} \bibinfo{person}{Marco~Tulio Valente}.} \bibinfo{year}{2018}\natexlab{}.
\newblock \showarticletitle{What’s in a github star? understanding repository starring practices in a social coding platform}.
\newblock \bibinfo{journal}{\emph{Journal of Systems and Software}}  \bibinfo{volume}{146} (\bibinfo{year}{2018}), \bibinfo{pages}{112--129}.
\newblock


\bibitem[Bosu and Sultana(2019)]%
        {bosu2019diversity}
\bibfield{author}{\bibinfo{person}{Amiangshu Bosu} {and} \bibinfo{person}{Kazi~Zakia Sultana}.} \bibinfo{year}{2019}\natexlab{}.
\newblock \showarticletitle{Diversity and inclusion in open source software (OSS) projects: Where do we stand?}. In \bibinfo{booktitle}{\emph{2019 ACM/IEEE International Symposium on Empirical Software Engineering and Measurement (ESEM)}}. IEEE, \bibinfo{pages}{1--11}.
\newblock


\bibitem[Bregolin(2022)]%
        {bregolin2022communication}
\bibfield{author}{\bibinfo{person}{Jacopo Bregolin}.} \bibinfo{year}{2022}\natexlab{}.
\newblock \bibinfo{booktitle}{\emph{Communication effort and the cost of language: Evidence from stack overflow}}.
\newblock \bibinfo{type}{{T}echnical {R}eport}.
\newblock


\bibitem[Buzatu(2025)]%
        {buzatu2025data}
\bibfield{author}{\bibinfo{person}{Bogdan-Mihai Buzatu}.} \bibinfo{year}{2025}\natexlab{}.
\newblock \emph{\bibinfo{title}{Data hound: Analysing non-English data smells in large code datasets}}.
\newblock \bibinfo{thesistype}{Ph.\,D. Dissertation}. \bibinfo{school}{Delft University of Technology}.
\newblock


\bibitem[Dabbish et~al\mbox{.}(2012)]%
        {dabbish2012social}
\bibfield{author}{\bibinfo{person}{Laura Dabbish}, \bibinfo{person}{Colleen Stuart}, \bibinfo{person}{Jason Tsay}, {and} \bibinfo{person}{Jim Herbsleb}.} \bibinfo{year}{2012}\natexlab{}.
\newblock \showarticletitle{Social coding in GitHub: transparency and collaboration in an open software repository}. In \bibinfo{booktitle}{\emph{Proceedings of the ACM 2012 conference on computer supported cooperative work}}. \bibinfo{pages}{1277--1286}.
\newblock


\bibitem[Daigle and {GitHub Staff}(2023)]%
        {githubOctoverseState2023}
\bibfield{author}{\bibinfo{person}{Kyle Daigle} {and} \bibinfo{person}{{GitHub Staff}}.} \bibinfo{year}{2023}\natexlab{}.
\newblock \bibinfo{title}{{Octoverse: The State of Open Source and Rise of AI in 2023 --- github.blog}}.
\newblock \bibinfo{howpublished}{\url{https://github.blog/news-insights/research/the-state-of-open-source-and-ai/}}.
\newblock


\bibitem[Denny et~al\mbox{.}(2021)]%
        {denny2021designing}
\bibfield{author}{\bibinfo{person}{Paul Denny}, \bibinfo{person}{James Prather}, \bibinfo{person}{Brett~A Becker}, \bibinfo{person}{Catherine Mooney}, \bibinfo{person}{John Homer}, \bibinfo{person}{Zachary~C Albrecht}, {and} \bibinfo{person}{Garrett~B Powell}.} \bibinfo{year}{2021}\natexlab{}.
\newblock \showarticletitle{On designing programming error messages for novices: Readability and its constituent factors}. In \bibinfo{booktitle}{\emph{Proceedings of the 2021 CHI Conference on Human Factors in Computing Systems}}. \bibinfo{pages}{1--15}.
\newblock


\bibitem[Destefanis et~al\mbox{.}(2018)]%
        {destefanis2018measuring}
\bibfield{author}{\bibinfo{person}{Giuseppe Destefanis}, \bibinfo{person}{Marco Ortu}, \bibinfo{person}{David Bowes}, \bibinfo{person}{Michele Marchesi}, {and} \bibinfo{person}{Roberto Tonelli}.} \bibinfo{year}{2018}\natexlab{}.
\newblock \showarticletitle{On measuring affects of github issues' commenters}. In \bibinfo{booktitle}{\emph{Proceedings of the 3rd International Workshop on Emotion Awareness in Software Engineering}}. \bibinfo{pages}{14--19}.
\newblock


\bibitem[Ebbertz(2002)]%
        {ebbertz2002internet}
\bibfield{author}{\bibinfo{person}{M Ebbertz}.} \bibinfo{year}{2002}\natexlab{}.
\newblock \bibinfo{title}{Internet statistics: Distribution of languages on the Internet}.
\newblock


\bibitem[{Eclipse Foundation}(2006)]%
        {eclipseEclipseCommunity}
\bibfield{author}{\bibinfo{person}{{Eclipse Foundation}}.} \bibinfo{year}{2006}\natexlab{}.
\newblock \bibinfo{title}{{Eclipse Community Forums: Eclipse Platform » Character Encoding Problem (on Windows) | The Eclipse Foundation --- eclipse.org}}.
\newblock \bibinfo{howpublished}{\url{https://www.eclipse.org/forums/index.php/t/104105/}}.
\newblock


\bibitem[Fatima(2025)]%
        {fatima2025developer}
\bibfield{author}{\bibinfo{person}{Urooj Fatima}.} \bibinfo{year}{2025}\natexlab{}.
\newblock \showarticletitle{Developer social networks/open source project networks: how programmers use GitHub}.
\newblock  (\bibinfo{year}{2025}).
\newblock


\bibitem[Feng et~al\mbox{.}(2025)]%
        {feng2025multifaceted}
\bibfield{author}{\bibinfo{person}{Zixuan Feng}, \bibinfo{person}{Igor Steinmacher}, \bibinfo{person}{Marco Gerosa}, \bibinfo{person}{Tyler Menezes}, \bibinfo{person}{Alexander Serebrenik}, \bibinfo{person}{Reed Milewicz}, {and} \bibinfo{person}{Anita Sarma}.} \bibinfo{year}{2025}\natexlab{}.
\newblock \showarticletitle{The multifaceted nature of mentoring in oss: strategies, qualities, and ideal outcomes}. In \bibinfo{booktitle}{\emph{2025 IEEE/ACM 18th International Conference on Cooperative and Human Aspects of Software Engineering (CHASE)}}. IEEE, \bibinfo{pages}{203--214}.
\newblock


\bibitem[{GitHub}(2021)]%
        {githubGitHubCopilot}
\bibfield{author}{\bibinfo{person}{{GitHub}}.} \bibinfo{year}{2021}\natexlab{}.
\newblock \bibinfo{title}{{GitHub Copilot · Your AI Pair Programmer --- github.com}}.
\newblock \bibinfo{howpublished}{\url{https://github.com/features/copilot}}.
\newblock


\bibitem[{GitHub}(2022)]%
        {githubGlobalDistribution2022}
\bibfield{author}{\bibinfo{person}{{GitHub}}.} \bibinfo{year}{2022}\natexlab{}.
\newblock \bibinfo{title}{{Global Distribution of Developers --- octoverse.github.com}}.
\newblock \bibinfo{howpublished}{\url{https://octoverse.github.com/2022/global-tech-talent}}.
\newblock


\bibitem[{GitHub Staff}(2024)]%
        {githubOctoverseLeads2024}
\bibfield{author}{\bibinfo{person}{{GitHub Staff}}.} \bibinfo{year}{2024}\natexlab{}.
\newblock \bibinfo{title}{{Octoverse: AI Leads Python to Top Language as the Number of Global Developers Surges --- github.blog}}.
\newblock \bibinfo{howpublished}{\url{https://github.blog/news-insights/octoverse/octoverse-2024/\#the-state-of-open-source}}.
\newblock


\bibitem[{GNU Project}(2016)]%
        {gnuWhenFree}
\bibfield{author}{\bibinfo{person}{{GNU Project}}.} \bibinfo{year}{2016}\natexlab{}.
\newblock \bibinfo{title}{{When Free Software Isn't (Practically) Superior --- GNU Project --- Free Software Foundation}}.
\newblock \bibinfo{howpublished}{\url{https://www.gnu.org/philosophy/when-free-software-isnt-practically-superior.html}}.
\newblock


\bibitem[Golzadeh et~al\mbox{.}(2021a)]%
        {golzadeh2021identifying}
\bibfield{author}{\bibinfo{person}{Mehdi Golzadeh}, \bibinfo{person}{Alexandre Decan}, \bibinfo{person}{Eleni Constantinou}, {and} \bibinfo{person}{Tom Mens}.} \bibinfo{year}{2021}\natexlab{a}.
\newblock \showarticletitle{Identifying bot activity in github pull request and issue comments}. In \bibinfo{booktitle}{\emph{2021 IEEE/ACM third international workshop on bots in software engineering (BotSE)}}. IEEE, \bibinfo{pages}{21--25}.
\newblock


\bibitem[Golzadeh et~al\mbox{.}(2021b)]%
        {golzadeh2021ground}
\bibfield{author}{\bibinfo{person}{Mehdi Golzadeh}, \bibinfo{person}{Alexandre Decan}, \bibinfo{person}{Damien Legay}, {and} \bibinfo{person}{Tom Mens}.} \bibinfo{year}{2021}\natexlab{b}.
\newblock \showarticletitle{A ground-truth dataset and classification model for detecting bots in GitHub issue and PR comments}.
\newblock \bibinfo{journal}{\emph{Journal of Systems and Software}}  \bibinfo{volume}{175} (\bibinfo{year}{2021}), \bibinfo{pages}{110911}.
\newblock


\bibitem[Golzadeh et~al\mbox{.}(2020)]%
        {golzadeh2020bot}
\bibfield{author}{\bibinfo{person}{Mehdi Golzadeh}, \bibinfo{person}{Damien Legay}, \bibinfo{person}{Alexandre Decan}, {and} \bibinfo{person}{Tom Mens}.} \bibinfo{year}{2020}\natexlab{}.
\newblock \showarticletitle{Bot or not? Detecting bots in GitHub pull request activity based on comment similarity}. In \bibinfo{booktitle}{\emph{Proceedings of the IEEE/ACM 42nd international conference on software engineering workshops}}. \bibinfo{pages}{31--35}.
\newblock


\bibitem[Gousios et~al\mbox{.}(2014)]%
        {gousios2014exploratory}
\bibfield{author}{\bibinfo{person}{Georgios Gousios}, \bibinfo{person}{Martin Pinzger}, {and} \bibinfo{person}{Arie~van Deursen}.} \bibinfo{year}{2014}\natexlab{}.
\newblock \showarticletitle{An exploratory study of the pull-based software development model}. In \bibinfo{booktitle}{\emph{Proceedings of the 36th international conference on software engineering}}. \bibinfo{pages}{345--355}.
\newblock


\bibitem[Guo(2014)]%
        {acmPythonMost}
\bibfield{author}{\bibinfo{person}{Philip Guo}.} \bibinfo{year}{2014}\natexlab{}.
\newblock \bibinfo{title}{{Python Is Now the Most Popular Introductory Teaching Language at Top U.S. Universities --- Communications of the ACM}}.
\newblock \bibinfo{howpublished}{\url{https://cacm.acm.org/blogcacm/python-is-now-the-most-popular-introductory-teaching-language-at-top-u-s-universities/}}.
\newblock


\bibitem[Guo(2018)]%
        {guo2018non}
\bibfield{author}{\bibinfo{person}{Philip~J Guo}.} \bibinfo{year}{2018}\natexlab{}.
\newblock \showarticletitle{Non-native english speakers learning computer programming: Barriers, desires, and design opportunities}. In \bibinfo{booktitle}{\emph{Proceedings of the 2018 CHI conference on human factors in computing systems}}. \bibinfo{pages}{1--14}.
\newblock


\bibitem[Hasan et~al\mbox{.}(2023)]%
        {hasan2023understandingtimeresponsegithub}
\bibfield{author}{\bibinfo{person}{Kazi~Amit Hasan}, \bibinfo{person}{Marcos Macedo}, \bibinfo{person}{Yuan Tian}, \bibinfo{person}{Bram Adams}, {and} \bibinfo{person}{Steven Ding}.} \bibinfo{year}{2023}\natexlab{}.
\newblock \showarticletitle{Understanding the time to first response in GitHub pull requests}. In \bibinfo{booktitle}{\emph{2023 IEEE/ACM 20th International Conference on Mining Software Repositories (MSR)}}. IEEE, \bibinfo{pages}{1--11}.
\newblock


\bibitem[He et~al\mbox{.}(2024)]%
        {he20244}
\bibfield{author}{\bibinfo{person}{Hao He}, \bibinfo{person}{Haoqin Yang}, \bibinfo{person}{Philipp Burckhardt}, \bibinfo{person}{Alexandros Kapravelos}, \bibinfo{person}{Bogdan Vasilescu}, {and} \bibinfo{person}{Christian K{\"a}stner}.} \bibinfo{year}{2024}\natexlab{}.
\newblock \showarticletitle{4.5 Million (Suspected) Fake Stars in GitHub: A Growing Spiral of Popularity Contests, Scams, and Malware}.
\newblock \bibinfo{journal}{\emph{arXiv preprint arXiv:2412.13459}} (\bibinfo{year}{2024}).
\newblock


\bibitem[Hellman et~al\mbox{.}(2022)]%
        {hellman2022characterizing}
\bibfield{author}{\bibinfo{person}{Jazlyn Hellman}, \bibinfo{person}{Jiahao Chen}, \bibinfo{person}{Md~Sami Uddin}, \bibinfo{person}{Jinghui Cheng}, {and} \bibinfo{person}{Jin~LC Guo}.} \bibinfo{year}{2022}\natexlab{}.
\newblock \showarticletitle{Characterizing user behaviors in open-source software user forums: an empirical study}. In \bibinfo{booktitle}{\emph{Proceedings of the 15th International Conference on Cooperative and Human Aspects of Software Engineering}}. \bibinfo{pages}{46--55}.
\newblock


\bibitem[Hindle et~al\mbox{.}(2008)]%
        {hindle2008large}
\bibfield{author}{\bibinfo{person}{Abram Hindle}, \bibinfo{person}{Daniel~M German}, {and} \bibinfo{person}{Ric Holt}.} \bibinfo{year}{2008}\natexlab{}.
\newblock \showarticletitle{What do large commits tell us? a taxonomical study of large commits}. In \bibinfo{booktitle}{\emph{Proceedings of the 2008 international working conference on Mining software repositories}}. \bibinfo{pages}{99--108}.
\newblock


\bibitem[Hyrynsalmi et~al\mbox{.}(2024)]%
        {hyrynsalmi2024bridging}
\bibfield{author}{\bibinfo{person}{Sonja~M Hyrynsalmi}, \bibinfo{person}{Sebastian Baltes}, \bibinfo{person}{Chris Brown}, \bibinfo{person}{Rafael Prikladnicki}, \bibinfo{person}{Gema Rodriguez-Perez}, \bibinfo{person}{Alexander Serebrenik}, \bibinfo{person}{Jocelyn Simmonds}, \bibinfo{person}{Bianca Trinkenreich}, \bibinfo{person}{Yi Wang}, {and} \bibinfo{person}{Grischa Liebel}.} \bibinfo{year}{2024}\natexlab{}.
\newblock \showarticletitle{Bridging gaps, building futures: Advancing software developer diversity and inclusion through future-oriented research}.
\newblock \bibinfo{journal}{\emph{arXiv preprint arXiv:2404.07142}} (\bibinfo{year}{2024}).
\newblock


\bibitem[{Interoperable Europe}(2022)]%
        {eu2022koreaoss}
\bibfield{author}{\bibinfo{person}{{Interoperable Europe}}.} \bibinfo{year}{2022}\natexlab{}.
\newblock \bibinfo{booktitle}{\emph{Open Source Software Country Intelligence Report South Korea}}.
\newblock \bibinfo{type}{{T}echnical {R}eport}. \bibinfo{institution}{European Commission}.
\newblock
\urldef\tempurl%
\url{https://interoperable-europe.ec.europa.eu/sites/default/files/inline-files/OSS%20Country%20Intelligence%20report_KR.pdf}
\showURL{%
\tempurl}


\bibitem[{JetBrains}(2018)]%
        {jetbrainsIntelliJFile}
\bibfield{author}{\bibinfo{person}{{JetBrains}}.} \bibinfo{year}{2018}\natexlab{}.
\newblock \bibinfo{title}{{IntelliJ File Encoding Issue --- intellij-support.jetbrains.com}}.
\newblock \bibinfo{howpublished}{\url{https://intellij-support.jetbrains.com/hc/en-us/community/posts/360003953579-IntelliJ-file-encoding-issue}}.
\newblock


\bibitem[{JetBrains}(2021)]%
        {jetbrainsJavaProgramming}
\bibfield{author}{\bibinfo{person}{{JetBrains}}.} \bibinfo{year}{2021}\natexlab{}.
\newblock \bibinfo{title}{{Java Programming --- The State of Developer Ecosystem in 2021 Infographic}}.
\newblock \bibinfo{howpublished}{\url{https://www.jetbrains.com/lp/devecosystem-2021/java/}}.
\newblock


\bibitem[Katzy et~al\mbox{.}(2025)]%
        {katzy2025qualitativeinvestigationllmgeneratedmultilingual}
\bibfield{author}{\bibinfo{person}{Jonathan Katzy}, \bibinfo{person}{Yongcheng Huang}, \bibinfo{person}{Gopal-Raj Panchu}, \bibinfo{person}{Maksym Ziemlewski}, \bibinfo{person}{Paris Loizides}, \bibinfo{person}{Sander Vermeulen}, \bibinfo{person}{Arie van Deursen}, {and} \bibinfo{person}{Maliheh Izadi}.} \bibinfo{year}{2025}\natexlab{}.
\newblock \showarticletitle{A Qualitative Investigation into LLM-Generated Multilingual Code Comments and Automatic Evaluation Metrics}.
\newblock  (\bibinfo{year}{2025}).
\newblock


\bibitem[Kavaler et~al\mbox{.}(2017)]%
        {kavaler2017perceived}
\bibfield{author}{\bibinfo{person}{David Kavaler}, \bibinfo{person}{Sasha Sirovica}, \bibinfo{person}{Vincent Hellendoorn}, \bibinfo{person}{Raul Aranovich}, {and} \bibinfo{person}{Vladimir Filkov}.} \bibinfo{year}{2017}\natexlab{}.
\newblock \showarticletitle{Perceived language complexity in GitHub issue discussions and their effect on issue resolution}. In \bibinfo{booktitle}{\emph{2017 32nd IEEE/ACM International Conference on Automated Software Engineering (ASE)}}. IEEE, \bibinfo{pages}{72--83}.
\newblock


\bibitem[Koch et~al\mbox{.}(2024)]%
        {koch2024fault}
\bibfield{author}{\bibinfo{person}{Simon Koch}, \bibinfo{person}{David Klein}, {and} \bibinfo{person}{Martin Johns}.} \bibinfo{year}{2024}\natexlab{}.
\newblock \showarticletitle{The Fault in Our Stars: An Analysis of GitHub Stars as an Importance Metric for Web Source Code}. In \bibinfo{booktitle}{\emph{Workshop on Measurements, Attacks, and Defenses for the Web (MADWeb)}}, Vol.~\bibinfo{volume}{2024}.
\newblock


\bibitem[Kondo et~al\mbox{.}(2024)]%
        {kondo2024empirical}
\bibfield{author}{\bibinfo{person}{Masanari Kondo}, \bibinfo{person}{Daniel~M German}, \bibinfo{person}{Yasutaka Kamei}, \bibinfo{person}{Naoyasu Ubayashi}, {and} \bibinfo{person}{Osamu Mizuno}.} \bibinfo{year}{2024}\natexlab{}.
\newblock \showarticletitle{An empirical study of token-based micro commits}.
\newblock \bibinfo{journal}{\emph{Empirical Software Engineering}} \bibinfo{volume}{29}, \bibinfo{number}{6} (\bibinfo{year}{2024}), \bibinfo{pages}{148}.
\newblock


\bibitem[Kosti{\'c} et~al\mbox{.}(2023)]%
        {kostic2023monolingual}
\bibfield{author}{\bibinfo{person}{Marija Kosti{\'c}}, \bibinfo{person}{Vuk Batanovi{\'c}}, {and} \bibinfo{person}{Bo{\v{s}}ko Nikoli{\'c}}.} \bibinfo{year}{2023}\natexlab{}.
\newblock \showarticletitle{Monolingual, multilingual and cross-lingual code comment classification}.
\newblock \bibinfo{journal}{\emph{Engineering Applications of Artificial Intelligence}}  \bibinfo{volume}{124} (\bibinfo{year}{2023}), \bibinfo{pages}{106485}.
\newblock


\bibitem[Laurent(2025)]%
        {laurentRATPStatus}
\bibfield{author}{\bibinfo{person}{Vincent Laurent}.} \bibinfo{year}{2025}\natexlab{}.
\newblock \bibinfo{title}{ratpstatus}.
\newblock \bibinfo{howpublished}{\url{https://github.com/wincelau/ratpstatus}}.
\newblock


\bibitem[Liao and Singh(2019)]%
        {githubBanSanctioned2019}
\bibfield{author}{\bibinfo{person}{Rita Liao} {and} \bibinfo{person}{Manish Singh}.} \bibinfo{year}{2019}\natexlab{}.
\newblock \bibinfo{title}{GitHub confirms it has blocked developers in Iran, Syria and Crimea}.
\newblock \bibinfo{howpublished}{\url{https://techcrunch.com/2019/07/29/github-ban-sanctioned-countries/}}.
\newblock


\bibitem[Mahoney(2005)]%
        {mahoney2005histories}
\bibfield{author}{\bibinfo{person}{Michael~S Mahoney}.} \bibinfo{year}{2005}\natexlab{}.
\newblock \showarticletitle{The histories of computing (s)}.
\newblock \bibinfo{journal}{\emph{Interdisciplinary science reviews}} \bibinfo{volume}{30}, \bibinfo{number}{2} (\bibinfo{year}{2005}), \bibinfo{pages}{119--135}.
\newblock


\bibitem[{Microsoft}(2023)]%
        {csDevKitFAQ}
\bibfield{author}{\bibinfo{person}{{Microsoft}}.} \bibinfo{year}{2023}\natexlab{}.
\newblock \bibinfo{title}{{C\# Dev Kit FAQ --- code.visualstudio.com}}.
\newblock \bibinfo{howpublished}{\url{https://code.visualstudio.com/docs/csharp/cs-dev-kit-faq\#_who-can-use-chash-dev-kit}}.
\newblock


\bibitem[Miller et~al\mbox{.}(2022)]%
        {miller2022did}
\bibfield{author}{\bibinfo{person}{Courtney Miller}, \bibinfo{person}{Sophie Cohen}, \bibinfo{person}{Daniel Klug}, \bibinfo{person}{Bogdan Vasilescu}, {and} \bibinfo{person}{Christian KaUstner}.} \bibinfo{year}{2022}\natexlab{}.
\newblock \showarticletitle{" Did you miss my comment or what?" understanding toxicity in open source discussions}. In \bibinfo{booktitle}{\emph{Proceedings of the 44th international conference on software engineering}}. \bibinfo{pages}{710--722}.
\newblock


\bibitem[Murgia et~al\mbox{.}(2018)]%
        {murgia2018exploratory}
\bibfield{author}{\bibinfo{person}{Alessandro Murgia}, \bibinfo{person}{Marco Ortu}, \bibinfo{person}{Parastou Tourani}, \bibinfo{person}{Bram Adams}, {and} \bibinfo{person}{Serge Demeyer}.} \bibinfo{year}{2018}\natexlab{}.
\newblock \showarticletitle{An exploratory qualitative and quantitative analysis of emotions in issue report comments of open source systems}.
\newblock \bibinfo{journal}{\emph{Empirical Software Engineering}} \bibinfo{volume}{23}, \bibinfo{number}{1} (\bibinfo{year}{2018}), \bibinfo{pages}{521--564}.
\newblock


\bibitem[Murphy et~al\mbox{.}(2016)]%
        {murphy2016analysis}
\bibfield{author}{\bibinfo{person}{Ellen Murphy}, \bibinfo{person}{Tom Crick}, {and} \bibinfo{person}{James~H Davenport}.} \bibinfo{year}{2016}\natexlab{}.
\newblock \showarticletitle{An analysis of introductory programming courses at UK universities}.
\newblock \bibinfo{journal}{\emph{arXiv preprint arXiv:1609.06622}} (\bibinfo{year}{2016}).
\newblock


\bibitem[{N-iX}(2023)]%
        {nixWhereHire}
\bibfield{author}{\bibinfo{person}{{N-iX}}.} \bibinfo{year}{2023}\natexlab{}.
\newblock \bibinfo{title}{{Where to Hire Java Developers: UK vs Ukraine vs India --- N-iX}}.
\newblock \bibinfo{howpublished}{\url{https://www.n-ix.com/where-hire-java-developers-uk-ukraine-india/}}.
\newblock


\bibitem[Noll et~al\mbox{.}(2011)]%
        {noll2011global}
\bibfield{author}{\bibinfo{person}{John Noll}, \bibinfo{person}{Sarah Beecham}, {and} \bibinfo{person}{Ita Richardson}.} \bibinfo{year}{2011}\natexlab{}.
\newblock \showarticletitle{Global software development and collaboration: barriers and solutions}.
\newblock \bibinfo{journal}{\emph{ACM inroads}} \bibinfo{volume}{1}, \bibinfo{number}{3} (\bibinfo{year}{2011}), \bibinfo{pages}{66--78}.
\newblock


\bibitem[{OpenAI}(2022)]%
        {chatgptChatGPT}
\bibfield{author}{\bibinfo{person}{{OpenAI}}.} \bibinfo{year}{2022}\natexlab{}.
\newblock \bibinfo{title}{{ChatGPT --- chatgpt.com}}.
\newblock \bibinfo{howpublished}{\url{https://chatgpt.com/}}.
\newblock


\bibitem[{OpenJDK}(2020)]%
        {openjdkLoading}
\bibfield{author}{\bibinfo{person}{{OpenJDK}}.} \bibinfo{year}{2020}\natexlab{}.
\newblock \bibinfo{title}{{JDK-8242257: Encoding Issue in Windows Terminal --- bugs.openjdk.org}}.
\newblock \bibinfo{howpublished}{\url{https://bugs.openjdk.org/browse/JDK-8242257?page=com.atlassian.jira.plugin.system.issuetabpanels\%3Aall-tabpanel}}.
\newblock


\bibitem[{Oracle}(2015)]%
        {oracleJavadoc}
\bibfield{author}{\bibinfo{person}{{Oracle}}.} \bibinfo{year}{2015}\natexlab{}.
\newblock \bibinfo{title}{{javadoc --- docs.oracle.com}}.
\newblock \bibinfo{howpublished}{\url{https://docs.oracle.com/javase/8/docs/technotes/tools/windows/javadoc.html}}.
\newblock


\bibitem[Ortu et~al\mbox{.}(2015)]%
        {ortu2015would}
\bibfield{author}{\bibinfo{person}{Marco Ortu}, \bibinfo{person}{Giuseppe Destefanis}, \bibinfo{person}{Mohamad Kassab}, \bibinfo{person}{Steve Counsell}, \bibinfo{person}{Michele Marchesi}, {and} \bibinfo{person}{Roberto Tonelli}.} \bibinfo{year}{2015}\natexlab{}.
\newblock \showarticletitle{Would you mind fixing this issue? an empirical analysis of politeness and attractiveness in software developed using agile boards}. In \bibinfo{booktitle}{\emph{International conference on Agile software development}}. Springer, \bibinfo{pages}{129--140}.
\newblock


\bibitem[Park et~al\mbox{.}(2025)]%
        {park2025detection}
\bibfield{author}{\bibinfo{person}{Shinwoo Park}, \bibinfo{person}{Hyundong Jin}, \bibinfo{person}{Jeong-won Cha}, {and} \bibinfo{person}{Yo-Sub Han}.} \bibinfo{year}{2025}\natexlab{}.
\newblock \showarticletitle{Detection of llm-paraphrased code and identification of the responsible llm using coding style features}.
\newblock \bibinfo{journal}{\emph{arXiv preprint arXiv:2502.17749}} (\bibinfo{year}{2025}).
\newblock


\bibitem[Pawelka and Juergens(2015)]%
        {pawelka2015code}
\bibfield{author}{\bibinfo{person}{Timo Pawelka} {and} \bibinfo{person}{Elmar Juergens}.} \bibinfo{year}{2015}\natexlab{}.
\newblock \showarticletitle{Is this code written in English? A study of the natural language of comments and identifiers in practice}. In \bibinfo{booktitle}{\emph{2015 IEEE International Conference on Software Maintenance and Evolution (ICSME)}}. IEEE, \bibinfo{pages}{401--410}.
\newblock


\bibitem[Pemistahl(2024)]%
        {lingua}
\bibfield{author}{\bibinfo{person}{Pemistahl}.} \bibinfo{year}{2024}\natexlab{}.
\newblock \bibinfo{title}{Lingua: An Accurate Natural Language Detection Library for Python}.
\newblock \bibinfo{howpublished}{\url{https://github.com/pemistahl/lingua-py}}.
\newblock


\bibitem[Prikladnicki et~al\mbox{.}(2003)]%
        {prikladnicki2003global}
\bibfield{author}{\bibinfo{person}{Rafael Prikladnicki}, \bibinfo{person}{Jorge~Luis Nicolas~Audy}, {and} \bibinfo{person}{Roberto Evaristo}.} \bibinfo{year}{2003}\natexlab{}.
\newblock \showarticletitle{Global software development in practice lessons learned}.
\newblock \bibinfo{journal}{\emph{Software Process: Improvement and Practice}} \bibinfo{volume}{8}, \bibinfo{number}{4} (\bibinfo{year}{2003}), \bibinfo{pages}{267--281}.
\newblock


\bibitem[{Python Software Foundation}(2001)]%
        {pythonDocstringConventions}
\bibfield{author}{\bibinfo{person}{{Python Software Foundation}}.} \bibinfo{year}{2001}\natexlab{}.
\newblock \bibinfo{title}{{PEP 257 – Docstring Conventions | peps.python.org}}.
\newblock \bibinfo{howpublished}{\url{https://peps.python.org/pep-0257/}}.
\newblock


\bibitem[Rauschmayer(2016)]%
        {exploringjsUnicode}
\bibfield{author}{\bibinfo{person}{Axel Rauschmayer}.} \bibinfo{year}{2016}\natexlab{}.
\newblock \bibinfo{title}{{26. Unicode in ES6 --- exploringjs.com}}.
\newblock \bibinfo{howpublished}{\url{https://exploringjs.com/es6/ch_unicode.html}}.
\newblock


\bibitem[Ray et~al\mbox{.}(2014)]%
        {ray2014large}
\bibfield{author}{\bibinfo{person}{Baishakhi Ray}, \bibinfo{person}{Daryl Posnett}, \bibinfo{person}{Vladimir Filkov}, {and} \bibinfo{person}{Premkumar Devanbu}.} \bibinfo{year}{2014}\natexlab{}.
\newblock \showarticletitle{A large scale study of programming languages and code quality in github}. In \bibinfo{booktitle}{\emph{Proceedings of the 22nd ACM SIGSOFT international symposium on foundations of software engineering}}. \bibinfo{pages}{155--165}.
\newblock


\bibitem[Reestman and Dorn(2019)]%
        {reestman2019native}
\bibfield{author}{\bibinfo{person}{Kyle Reestman} {and} \bibinfo{person}{Brian Dorn}.} \bibinfo{year}{2019}\natexlab{}.
\newblock \showarticletitle{Native language's effect on java compiler errors}. In \bibinfo{booktitle}{\emph{Proceedings of the 2019 ACM conference on international computing education research}}. \bibinfo{pages}{249--257}.
\newblock


\bibitem[Santhanam et~al\mbox{.}(2022)]%
        {santhanam2022bots}
\bibfield{author}{\bibinfo{person}{Sivasurya Santhanam}, \bibinfo{person}{Tobias Hecking}, \bibinfo{person}{Andreas Schreiber}, {and} \bibinfo{person}{Stefan Wagner}.} \bibinfo{year}{2022}\natexlab{}.
\newblock \showarticletitle{Bots in software engineering: a systematic mapping study}.
\newblock \bibinfo{journal}{\emph{PeerJ Computer Science}}  \bibinfo{volume}{8} (\bibinfo{year}{2022}), \bibinfo{pages}{e866}.
\newblock


\bibitem[Santos et~al\mbox{.}(2025)]%
        {santos2025understanding}
\bibfield{author}{\bibinfo{person}{Reydne Santos}, \bibinfo{person}{Rafa Prado}, \bibinfo{person}{Ana~Paula de Holanda~Silva}, \bibinfo{person}{Kiev Gama}, \bibinfo{person}{Fernando Castor}, {and} \bibinfo{person}{Ronnie de Souza~Santos}.} \bibinfo{year}{2025}\natexlab{}.
\newblock \showarticletitle{Understanding Underrepresented Groups in Open Source Software}. In \bibinfo{booktitle}{\emph{Proceedings of the 29th International Conference on Evaluation and Assessment in Software Engineering}}. \bibinfo{pages}{919--928}.
\newblock


\bibitem[Sarker et~al\mbox{.}(2023)]%
        {ToxiSpanSE}
\bibfield{author}{\bibinfo{person}{Jaydeb Sarker}, \bibinfo{person}{Sayma Sultana}, \bibinfo{person}{Steven~R. Wilson}, {and} \bibinfo{person}{Amiangshu Bosu}.} \bibinfo{year}{2023}\natexlab{}.
\newblock \showarticletitle{ToxiSpanSE: An Explainable Toxicity Detection in Code Review Comments}. In \bibinfo{booktitle}{\emph{2023 ACM/IEEE International Symposium on Empirical Software Engineering and Measurement (ESEM)}}. \bibinfo{pages}{1--12}.
\newblock
\href{https://doi.org/10.1109/ESEM56168.2023.10304855}{doi:\nolinkurl{10.1109/ESEM56168.2023.10304855}}


\bibitem[Sarker et~al\mbox{.}(2025)]%
        {landscapetoxic}
\bibfield{author}{\bibinfo{person}{Jaydeb Sarker}, \bibinfo{person}{Asif~Kamal Turzo}, {and} \bibinfo{person}{Amiangshu Bosu}.} \bibinfo{year}{2025}\natexlab{}.
\newblock \showarticletitle{The Landscape of Toxicity: An Empirical Investigation of Toxicity on GitHub}.
\newblock \bibinfo{journal}{\emph{Proc. ACM Softw. Eng.}} \bibinfo{volume}{2}, \bibinfo{number}{FSE}, Article \bibinfo{articleno}{FSE029} (\bibinfo{date}{June} \bibinfo{year}{2025}), \bibinfo{numpages}{24}~pages.
\newblock
\href{https://doi.org/10.1145/3715744}{doi:\nolinkurl{10.1145/3715744}}


\bibitem[{Stack Overflow}({[n.\,d.]})]%
        {stackoverflowStackOverflow}
\bibfield{author}{\bibinfo{person}{{Stack Overflow}}.} \bibinfo{year}{[n.\,d.]}\natexlab{}.
\newblock \bibinfo{title}{{S}tack {O}verflow --- survey.stackoverflow.co}.
\newblock \bibinfo{howpublished}{\url{https://survey.stackoverflow.co/}}.
\newblock


\bibitem[{Stack Overflow}(2024)]%
        {stackoverflowQuestionLanguage}
\bibfield{author}{\bibinfo{person}{{Stack Overflow}}.} \bibinfo{year}{2024}\natexlab{}.
\newblock \bibinfo{title}{{Can I Ask a Question in a Language Other Than English? --- Help Center}}.
\newblock \bibinfo{howpublished}{\url{https://stackoverflow.com/help/non-english-questions}}.
\newblock


\bibitem[Staff({[n.\,d.]})]%
        {octoverse2025}
\bibfield{author}{\bibinfo{person}{GitHub Staff}.} \bibinfo{year}{[n.\,d.]}\natexlab{}.
\newblock \bibinfo{title}{{O}ctoverse: {A} new developer joins {G}it{H}ub every second as {A}{I} leads {T}ype{S}cript to \#1 --- github.blog}.
\newblock \bibinfo{howpublished}{\url{https://github.blog/news-insights/octoverse/octoverse-a-new-developer-joins-github-every-second-as-ai-leads-typescript-to-1/\#h-the-state-of-github-in-2025-a-year-of-record-growth}}.
\newblock


\bibitem[Steinmacher et~al\mbox{.}(2015a)]%
        {steinmacher2015social}
\bibfield{author}{\bibinfo{person}{Igor Steinmacher}, \bibinfo{person}{Tayana Conte}, \bibinfo{person}{Marco~Aur{\'e}lio Gerosa}, {and} \bibinfo{person}{David Redmiles}.} \bibinfo{year}{2015}\natexlab{a}.
\newblock \showarticletitle{Social barriers faced by newcomers placing their first contribution in open source software projects}. In \bibinfo{booktitle}{\emph{Proceedings of the 18th ACM conference on Computer supported cooperative work \& social computing}}. \bibinfo{pages}{1379--1392}.
\newblock


\bibitem[Steinmacher et~al\mbox{.}(2019)]%
        {steinmacher2019overcoming}
\bibfield{author}{\bibinfo{person}{Igor Steinmacher}, \bibinfo{person}{Marco Gerosa}, \bibinfo{person}{Tayana~U Conte}, {and} \bibinfo{person}{David~F Redmiles}.} \bibinfo{year}{2019}\natexlab{}.
\newblock \showarticletitle{Overcoming social barriers when contributing to open source software projects}.
\newblock \bibinfo{journal}{\emph{Computer Supported Cooperative Work (CSCW)}} \bibinfo{volume}{28}, \bibinfo{number}{1} (\bibinfo{year}{2019}), \bibinfo{pages}{247--290}.
\newblock


\bibitem[Steinmacher et~al\mbox{.}(2015b)]%
        {steinmacher2015systematic}
\bibfield{author}{\bibinfo{person}{Igor Steinmacher}, \bibinfo{person}{Marco Aurelio~Graciotto Silva}, \bibinfo{person}{Marco~Aurelio Gerosa}, {and} \bibinfo{person}{David~F Redmiles}.} \bibinfo{year}{2015}\natexlab{b}.
\newblock \showarticletitle{A systematic literature review on the barriers faced by newcomers to open source software projects}.
\newblock \bibinfo{journal}{\emph{Information and Software Technology}}  \bibinfo{volume}{59} (\bibinfo{year}{2015}), \bibinfo{pages}{67--85}.
\newblock


\bibitem[Sultana et~al\mbox{.}(2024)]%
        {sultana2024assessinginfluencetoxicgender}
\bibfield{author}{\bibinfo{person}{Sayma Sultana}, \bibinfo{person}{Gias Uddin}, {and} \bibinfo{person}{Amiangshu Bosu}.} \bibinfo{year}{2024}\natexlab{}.
\newblock \showarticletitle{Assessing the influence of toxic and gender discriminatory communication on perceptible diversity in oss projects}.
\newblock \bibinfo{journal}{\emph{arXiv preprint arXiv:2403.08113}} (\bibinfo{year}{2024}).
\newblock


\bibitem[Terrell et~al\mbox{.}(2016)]%
        {terrell2016gender}
\bibfield{author}{\bibinfo{person}{J Terrell}, \bibinfo{person}{A Kofink}, \bibinfo{person}{J Middleton}, \bibinfo{person}{C Rainear}, \bibinfo{person}{E Murphy-Hill}, {and} \bibinfo{person}{C Parnin}.} \bibinfo{year}{2016}\natexlab{}.
\newblock \showarticletitle{Gender bias in open source: Pull request acceptance of women versus men.(Jan 2016)}.
\newblock \bibinfo{journal}{\emph{PeerJ Computer Science}} (\bibinfo{year}{2016}).
\newblock


\bibitem[{Tree-sitter Project}(2024)]%
        {treesitterIntroductionTreesitter}
\bibfield{author}{\bibinfo{person}{{Tree-sitter Project}}.} \bibinfo{year}{2024}\natexlab{}.
\newblock \bibinfo{title}{{Introduction --- Tree-sitter}}.
\newblock \bibinfo{howpublished}{\url{https://tree-sitter.github.io/tree-sitter/index.html}}.
\newblock


\bibitem[Twist et~al\mbox{.}(2025)]%
        {twist2025llmslovepythonstudy}
\bibfield{author}{\bibinfo{person}{Lukas Twist}, \bibinfo{person}{Jie~M Zhang}, \bibinfo{person}{Mark Harman}, \bibinfo{person}{Don Syme}, \bibinfo{person}{Joost Noppen}, {and} \bibinfo{person}{Detlef Nauck}.} \bibinfo{year}{2025}\natexlab{}.
\newblock \showarticletitle{LLMs Love Python: A Study of LLMs' Bias for Programming Languages and Libraries}.
\newblock \bibinfo{journal}{\emph{arXiv preprint arXiv:2503.17181}} (\bibinfo{year}{2025}).
\newblock


\bibitem[{Unicode Consortium}({[n.\,d.]})]%
        {unicodeUnicode}
\bibfield{author}{\bibinfo{person}{{Unicode Consortium}}.} \bibinfo{year}{[n.\,d.]}\natexlab{}.
\newblock \bibinfo{title}{{U}nicode 5.1.0 --- unicode.org}.
\newblock \bibinfo{howpublished}{\url{https://www.unicode.org/versions/Unicode5.1.0/}}.
\newblock


\bibitem[Vasilescu et~al\mbox{.}(2012)]%
        {genderrepresentation}
\bibfield{author}{\bibinfo{person}{Bogdan Vasilescu}, \bibinfo{person}{Andrea Capiluppi}, {and} \bibinfo{person}{Alexander Serebrenik}.} \bibinfo{year}{2012}\natexlab{}.
\newblock \showarticletitle{Gender, Representation and Online Participation: A Quantitative Study of StackOverflow}. In \bibinfo{booktitle}{\emph{2012 International Conference on Social Informatics}}. \bibinfo{pages}{332--338}.
\newblock
\href{https://doi.org/10.1109/SocialInformatics.2012.81}{doi:\nolinkurl{10.1109/SocialInformatics.2012.81}}


\bibitem[Vasilescu et~al\mbox{.}(2015)]%
        {vasilescu2015gender}
\bibfield{author}{\bibinfo{person}{Bogdan Vasilescu}, \bibinfo{person}{Daryl Posnett}, \bibinfo{person}{Baishakhi Ray}, \bibinfo{person}{Mark~GJ van~den Brand}, \bibinfo{person}{Alexander Serebrenik}, \bibinfo{person}{Premkumar Devanbu}, {and} \bibinfo{person}{Vladimir Filkov}.} \bibinfo{year}{2015}\natexlab{}.
\newblock \showarticletitle{Gender and tenure diversity in GitHub teams}. In \bibinfo{booktitle}{\emph{Proceedings of the 33rd annual ACM conference on human factors in computing systems}}. \bibinfo{pages}{3789--3798}.
\newblock


\bibitem[Wachs et~al\mbox{.}(2022)]%
        {wachs2022geography}
\bibfield{author}{\bibinfo{person}{Johannes Wachs}, \bibinfo{person}{Mariusz Nitecki}, \bibinfo{person}{William Schueller}, {and} \bibinfo{person}{Axel Polleres}.} \bibinfo{year}{2022}\natexlab{}.
\newblock \showarticletitle{The geography of open source software: evidence from github}.
\newblock \bibinfo{journal}{\emph{Technological Forecasting and Social Change}}  \bibinfo{volume}{176} (\bibinfo{year}{2022}), \bibinfo{pages}{121478}.
\newblock


\bibitem[Wang et~al\mbox{.}(2024)]%
        {wang2024uncovering}
\bibfield{author}{\bibinfo{person}{Yi Wang}, \bibinfo{person}{Yang Yue}, \bibinfo{person}{Wei Wang}, {and} \bibinfo{person}{Gaowei Zhang}.} \bibinfo{year}{2024}\natexlab{}.
\newblock \showarticletitle{Uncovering non-native speakers’ experiences in global software development teams——A bourdieusian perspective}.
\newblock \bibinfo{journal}{\emph{Computer Supported Cooperative Work (CSCW)}} (\bibinfo{year}{2024}), \bibinfo{pages}{1--36}.
\newblock


\bibitem[Wessel et~al\mbox{.}(2018)]%
        {wessel2018power}
\bibfield{author}{\bibinfo{person}{Mairieli Wessel}, \bibinfo{person}{Bruno~Mendes De~Souza}, \bibinfo{person}{Igor Steinmacher}, \bibinfo{person}{Igor~S Wiese}, \bibinfo{person}{Ivanilton Polato}, \bibinfo{person}{Ana~Paula Chaves}, {and} \bibinfo{person}{Marco~A Gerosa}.} \bibinfo{year}{2018}\natexlab{}.
\newblock \showarticletitle{The power of bots: Characterizing and understanding bots in oss projects}.
\newblock \bibinfo{journal}{\emph{Proceedings of the ACM on Human-Computer Interaction}} \bibinfo{volume}{2}, \bibinfo{number}{CSCW} (\bibinfo{year}{2018}), \bibinfo{pages}{1--19}.
\newblock


\bibitem[{Wikipedia}(2024a)]%
        {cplus}
\bibfield{author}{\bibinfo{person}{{Wikipedia}}.} \bibinfo{year}{2024}\natexlab{a}.
\newblock \bibinfo{title}{{C++23 --- Wikipedia}}.
\newblock \bibinfo{howpublished}{\url{https://en.wikipedia.org/wiki/C\%2B\%2B23}}.
\newblock


\bibitem[{Wikipedia}(2024b)]%
        {wikipediaTypeScriptWikipedia}
\bibfield{author}{\bibinfo{person}{{Wikipedia}}.} \bibinfo{year}{2024}\natexlab{b}.
\newblock \bibinfo{title}{{TypeScript --- Wikipedia}}.
\newblock \bibinfo{howpublished}{\url{https://en.wikipedia.org/wiki/TypeScript}}.
\newblock


\bibitem[{Wikipedia contributors}(2025)]%
        {githubCensorshipWiki}
\bibfield{author}{\bibinfo{person}{{Wikipedia contributors}}.} \bibinfo{year}{2025}\natexlab{}.
\newblock \bibinfo{title}{Censorship of GitHub}.
\newblock \bibinfo{howpublished}{\url{https://en.wikipedia.org/wiki/Censorship_of_GitHub}}.
\newblock


\bibitem[yuncaiji(2025)]%
        {yuncaijiAPI}
\bibfield{author}{\bibinfo{person}{yuncaiji}.} \bibinfo{year}{2025}\natexlab{}.
\newblock \bibinfo{title}{API}.
\newblock \bibinfo{howpublished}{\url{https://github.com/yuncaiji/API}}.
\newblock


\bibitem[Zhang et~al\mbox{.}(2013)]%
        {zhang2013predicting}
\bibfield{author}{\bibinfo{person}{Hongyu Zhang}, \bibinfo{person}{Liang Gong}, {and} \bibinfo{person}{Steve Versteeg}.} \bibinfo{year}{2013}\natexlab{}.
\newblock \showarticletitle{Predicting bug-fixing time: an empirical study of commercial software projects}. In \bibinfo{booktitle}{\emph{2013 35th International Conference on Software Engineering (ICSE)}}. IEEE, \bibinfo{pages}{1042--1051}.
\newblock


\bibitem[Zhang et~al\mbox{.}(2017)]%
        {zhang2017detecting}
\bibfield{author}{\bibinfo{person}{Yun Zhang}, \bibinfo{person}{David Lo}, \bibinfo{person}{Pavneet~Singh Kochhar}, \bibinfo{person}{Xin Xia}, \bibinfo{person}{Quanlai Li}, {and} \bibinfo{person}{Jianling Sun}.} \bibinfo{year}{2017}\natexlab{}.
\newblock \showarticletitle{Detecting similar repositories on GitHub}. In \bibinfo{booktitle}{\emph{2017 IEEE 24th International Conference on Software Analysis, Evolution and Reengineering (SANER)}}. IEEE, \bibinfo{pages}{13--23}.
\newblock


\bibitem[Zhao and Deek(2004)]%
        {zhao2004user}
\bibfield{author}{\bibinfo{person}{Luyin Zhao} {and} \bibinfo{person}{Fadi~P Deek}.} \bibinfo{year}{2004}\natexlab{}.
\newblock \showarticletitle{User collaboration in open source software development}.
\newblock \bibinfo{journal}{\emph{Electronic Markets}} \bibinfo{volume}{14}, \bibinfo{number}{2} (\bibinfo{year}{2004}), \bibinfo{pages}{89--103}.
\newblock


\end{thebibliography}

\end{document}